# Geo-Standardizing 3D Modeling of Surface/Subsurface Objects and Related Logical Spaces on Celestial Bodies: Case Studies for Moon and Mars

Dogus Guler[1,*], Demet Cilden-Guler[2]

[1] Schulich School of Engineering, University of Calgary, Calgary, Canada, dogus.guler@ucalgary.ca, ORCID: 0000-0002-3191-103X

[2] SPACE Lab, Department of Mechanical and Manufacturing Engineering, University of Calgary, Calgary, Canada, demet.cildenguler@ucalgary.ca, ORCID: 0000-0002-3924-5422

* Corresponding author

**Geo-Standardizing 3D Modeling of Surface/Subsurface Objects and Related Logical Spaces on Celestial Bodies: Case Studies for Moon and Mars**

## Abstract

Establishing frameworks for promoting the realization of various activities on celestial bodies sustainably is of great significance for different contexts, such as preserving the scientific evidence and space heritage. Therefore, this research first proposes a conceptual model that covers the different types of features, attributes, and relationships between them to comprehensively delineate the surface/subsurface objects and related logical spaces on celestial bodies. It then implements this conceptual model as a CityJSON extension in such a way that allows for creating the three-dimensional (3D) geodatasets that represent these objects and spaces in a standardized manner. Moreover, the usefulness of this study is demonstrated through creating CityJSON datasets that include 3D models of exemplary surface/subsurface objects from the Moon and Mars, such as a historical landing site and related logical spaces, such as exclusion zones for protecting this site. The results of the current study show that there is a strong potential for forming 3D geodatasets on celestial bodies that can provide a notable foundation for the technical implementation of international agreements and legal frameworks. This work also contributes to the design of planetary spatial data infrastructures (PSDI) by incorporating the third dimension.

## 1 Introduction

There is a growing interest in space projects to celestial bodies, particularly the Moon and Mars. The related technological developments regarding advancements in reusable space rocket boosters, such as SpaceX's Falcon (SpaceX, 2025) and recently SpaceX's Starship[1] and Blue Origin's New Glenn[2] assist these projects. While one of the important objectives within these types of space projects is scientific exploration, another one is related to mining activities that are aimed at valuable materials on the celestial bodies. The increased number of space-based start-ups and investments, and consequently the growth of the space economy, show evidence for this issue (ESA, 2025; World Economic Forum, 2024). For example, $7.8B was invested in this type of start-up in 2024, as stated in the Start-Up Space 2025 report (Bryce Tech, 2025). In this connection, there exist planned missions that target the Moon. For instance, while Artemis III aims for landing on the lunar surface soon (NASA, 2024), there is also an international partnership to establish the International Lunar Research Station (X. Wu, 2023). In addition, a recently announced executive order titled Ensuring American Space Superiority[3] includes the ambition for settling a permanent lunar outpost.

There are early international declarations and treaties where it is mentioned that the exploration and use of outer space should be carried out in the sense of benefit and interest for all humankind (UN, 1963, 1967). The Moon Agreement, which entered into force in 1984, also suggests the common heritage of mankind approach for the Moon and its natural resources (UN, 1979). However, this agreement was accepted by a limited number of countries. The interest in space mining consequently increased the volume of discussions on how space resources can be exploited. The discussions covered the Space Resource Exploration and Utilization Act of 2015[4] , which mentions assigning property rights to the entities that obtain the asteroid resources in outer space. After that, different countries such as Luxembourg and Japan prepared domestic laws that encompass giving property rights to their citizens regarding space resources (Depagter, 2022). The most recent international agreement is the Artemis Accords that aims to pave the way for enabling principles for exploration and use of celestial bodies (NASA, 2020).

---

[1] https://www.spacex.com/vehicles/starship

[2] https://www.blueorigin.com/missions/ng-2

[3] https://www.whitehouse.gov/presidential-actions/executive-orders/

[4] https://www.congress.gov/bill/114th-congress/house-bill/1508

As of 11 June 2026, 67 nations have signed this agreement. Another important resource with regard to establishing an international framework for space resource activities covers different building blocks in several contexts, such as priority and resource rights (COPUOS, 2020). The safety zone concept within the proposed building blocks was included in the Artemis Accords.

Another aspect regarding the celestial bodies, particularly the Moon, is space heritage. The discussions continue on how to preserve the heritage of mankind, such as historical landing sites on the Moon, with a similar mechanism to the World Heritage (Su & Li, 2025). In the sense of a legal framework, the mining areas and historical sites are within the topic of land administration on Earth, which manages the cadastral rights, restrictions, and responsibilities (RRR) regarding below and above the land surface (Williamson et al., 2010). Land administration systems provide the technical implementation as a special form of geographic information systems (GIS) where the spatial information and semantics are integrated, such as a spatial dataset that represents the land parcel with its attributes on property ownership (Guler & Yomralioglu, 2022). While these systems commonly encompass two-dimensional (2D) spatial datasets, they evolve into systems that can store and manage information in three-dimensional (3D) in order to sufficiently fulfil the requirements regarding cadastral RRR in the built environment (Kalogianni et al., 2020). Both managing and planning land subsurface are important factors that promote this evolvement because the 2D spatial datasets might be insufficient to unambiguously delineate the cadastral RRR below the land (Saeidian et al., 2023). Therefore, integrating 3D spatial datasets would be beneficial in the context of planning and managing mining activities on celestial bodies. In the sense of cadastral restriction, while one aspect is preserving the historical sites, another one is the protection of the areas on celestial bodies that are of scientific interest, such as water-ice locations on the Moon. Similar to the mining context, the third dimension should be considered within both subjects. One reason for this is that external effects that might harm the mentioned areas and sites cannot be depicted comprehensively in 2D (Spennemann & Murphy, 2020).

The second issue is that the subsurface of the areas having scientific value should also be restricted since there is a strong possibility that the scientific evidence lies underground. The important thing is that the mentioned restrictions and right of use/exploitation regarding mining activities can be delineated through 3D volumetric objects that represent logical spaces. Here, it can be noted that the use of physical and logical spaces is in line with the definition of bona fide and fiat objects by Smith & Varzi (2000). Physical spaces that correspond to bona fide objects can be defined as spaces that are bounded by physical boundaries/objects. Buildings and rooms that are physically bounded by slabs and walls are examples of physical space. However, there are logical spaces (i.e., fiat objects) that can be defined through virtual boundaries. City/district borders that are determined through virtual boundaries for administrative purposes or land parcels that are expressed by property lines are examples of logical spaces. In addition, logical spaces are utilized to delineate 3D objects that serve specific purposes, such as subsurface corridors that represent assigned rights of use for minerals. Therefore, the concept of logical spaces is widely used for both urban modeling (Kutzner et al., 2020) and land administration studies where they are also expressed as legal spaces (Guler, 2024; Kara et al., 2024; Saeidian et al., 2023).

Spatial data standardization that promotes the efficient use of spatial data resources is one of the essential issues. In this sense, standards regarding geodatasets are highly significant to enable data interoperability (Hjelmager et al., 2008). This issue is a vital counterpart of spatial data infrastructures (SDI) in the terrestrial context. In this connection, the increased amount of extraterrestrial-based spatial datasets and the lack of standardization on the metadata and semantics of these datasets resulted in the potential idea for establishing the planetary spatial data infrastructures (PSDI) (Hare et al., 2018; J. Laura et al., 2017). Considering the aforementioned needs for 3D representations regarding different contexts, such as preservation and land use planning, these infrastructures should have the capability of storing and sharing the 3D geodatasets. In light of this information, the main aim of the present study is to develop the geoinformation-based standard extension that allows for creating the 3D standardized geodatasets that represent the different surface/subsurface objects and related logical spaces on celestial bodies in an interoperable manner.

The paper continues with the background section that informs the reader regarding the literature on the related topics and data standards. This section also identifies the research gaps and, consequently, outlines the contributions of this study. Section 3 elaborates the methodological workflow of the study. Section 4 details how the conceptual model and corresponding standard extension are developed. Section 5 presents the demonstrations through exemplary cases by providing the details on data organization and data transformation. The last section discusses the results of the present study, as well as provides conclusions that can be used as a foundation for future studies.

# 2 Background

## 2.1 Overview of the Literature on Related Topics

In this section, the overview of the literature is presented regarding the topics that provide a foundation for the content of the current study. The first sub-section covers the property rights and interrelatedly space mining. Section 2.1.2 encompasses the space heritage context as another relevant topic in the sense of modeling surface/subsurface objects on celestial bodies. Lastly, Section 2.1.3 includes the planetary spatial datasets since the current work focuses on the spatial data modeling.

### 2.1.1 Property Rights and Space Mining

The discussion on legal background and, thereby, property rights, continues for quite a long time. For example, Pop (2000) discussed if/how land ownership can be applied in celestial bodies and mentioned that the United Nations (UN), where the different sovereigns are committed, can be the central organization to manage the extraterrestrial real estate for the benefit of all humans. Some authors discussed the existence of private property rights in the sense of possible appropriation of different areas of space, such as asteroids (Becerra, 2017). Gugunskiy et al. (2020) stated that the existing legal background might be insufficient to meet the needs of today's community on space exploration and utilisation of space resources. Similarly, Kostenko (2020) discussed the challenges for international space law in terms of current trends such as space tourism. While some authors underlined the inevitability of property rights (Tracz, 2023), some of them focused on the issue in the sense of global commons (Mei, 2024).

Apart from this, one of the related discussions from a legal perspective is the sustainability of space and its resources. Some researchers focused on this issue in the context of privatization of space exploration and stability of private activities (Iliopoulos & Esteban, 2020; Newman & Williamson, 2018; Solomon, 2017). In addition, legal issues regarding space mining and related technologies were discussed by various scholars (Anderson et al., 2023; Li, 2024; Svec, 2022; Xu, 2020). An additional sustainable development goal, "Space Environment", complementing existing ones, was also suggested in the sense of planetary sustainability (Galli & Losch, 2023). To maintain the sustainability of space, Gheorghe & Yuchnovicz (2015) proposed establishing the space infrastructure vulnerability cadastre.

The content of discussions started to cover national legal developments, such as the US Space Resource Exploration and Utilization Act that contains assigning possible property rights to entities on space resources that they obtained (Depagter, 2022; Tronchetti, 2015; von der Dunk, 2023). These discussions also encompassed the possible implications of international agreements such as the Artemis Accords (Deplano, 2021). For example, McKeown et al. (2022) discussed the practicability of the safety zone concept, which is included in the Artemis Accords, for commercial lunar mining activities by exemplifying the scenarios that consider different distances for safety zones. Sanders et al. (2023) mentioned the importance of the zone of permission for space mining activities. In addition, some authors discussed the benefit-sharing approaches regarding space mining by considering terrestrial approaches, such as deep-sea mining (Butkevičienė & Rabitz, 2022).

In light of these developments, the need for frameworks was expressed in the literature. While some authors focused on establishing a code of conduct (Chrysaki, 2020), some of them underlined the development of an international framework (Salmeri & Carlo, 2021). Recently, an international lunar resource campaign was also proposed (Neal et al., 2024). Additionally, scholars proposed off-Earth mining and environmental impact assessment frameworks for space resource extraction (Dallas et al., 2020, 2021). Steffen (2022) explained the data-driven framework that might be beneficial for space mining activities by promoting the sharing of the related data, such as locations of planned spacecraft. A few studies mentioned the land use policy and land management perspectives. For example, de Vries & Hugentobler (2021) discussed property rights by considering space debris under the land management concept. Whereas Dapremont (2021) expressed the views for possible land use policy approaches on Mars, the possible applicability of the cadastre concept in space was also discussed (Yomralioglu, 2024). Hubbard et al. (2024) proposed a mining code that benefits from a map-based approach that incorporates the management of mining activities using spatial planning.

In summary, a critical observation is that the existing research on property rights and space mining as relatedly marks the need for technical procedures that facilitate the implementation of a legal framework and support sustainable explorations on celestial bodies. Therefore, the present study aims to use the above-mentioned issues as a promoter for conceptualizing the 3D modeling standardization.

### 2.1.2 Space Heritage

Another topic discussed in the literature is related on how to maintain space heritage. In this sense, the concept of a planetary park that aims to define specific areas, such as heritage sites on celestial bodies, with the aim of protection, was proposed (C. Cockell & Horneck, 2004; C. S. Cockell, 2024). There is a policy paper that was published by NASA, which recommends the conservation of different historical sites through restriction buffer zones, such as landing sites of previous Moon missions (NASA, 2011). Walsh (2012) proposed a new protocol that promotes the protection of space heritage. Spennemann & Murphy (2020) discussed the historical sites on the Moon and suggested the implementation of exclusion zones. The authors also mentioned that determining these zones in 3D would be more beneficial because of the possible effects of spacecraft, such as rockets. In addition, international legal aspects for the protection of space heritage were discussed by the authors (Savelev & Khayrutdinov, 2020). Preparing the list of cultural heritage and protected sites was included in the Vancouver Recommendations on Space Mining (Outer Space Institute, 2020). Elvis et al. (2021) emphasized the importance of governance and justice implications by considering the scarcity of valuable sites on the Moon, including cultural sites. Perez-Alvaro (2024) expressed the value of heritage places in outer space in terms of extreme places by mentioning the similarities with underwater. Recently, Su & Li (2025) underlined the need for an international framework for space heritage.

A key point is that enabling data standardization would be significantly helpful for efforts regarding space heritage, similar to heritage on Earth. Especially, 3D spatial datasets require a vital consideration as seen in the world heritage context. In this manner, the current study aims to provide a comprehensive data modeling approach that promotes for successfully achieving space heritage recording/archiving and conservation.

### 2.1.3 Planetary Spatial Datasets

Noteworthy to mention is that there are large-volume datasets regarding planetary sciences, and this volume is significantly increasing because of the planned missions. These datasets cover a large type of data from different contexts such as atmosphere, geoscience, and navigation. For example, the size of the Planetary Data System (PDS) operated by NASA was reported as approximately 1 Petabyte (PB) in a 2017 report (NASA, 2017). In early 2026, the yearly data access and download usage ratios regarding the Planetary Science Division (PSD) that encompasses PDS are mentioned as 14,292 PB and 64 PB, respectively, and the volume of PSD is projected to 3,86 PB for 2030 (NASA, 2026). The increase in the volume of the planetary science-related datasets was similarly expressed by the European Space Agency (ESA) as well (Besse et al., 2018). Given the large size and various types of datasets that are obtained through space-based missions and extensive usage within planetary-related research, one topic of the discussions was related to how to achieve the efficient use of planetary spatial datasets. It can be noted that these types of datasets can be considered as covering topographical features and primitives on the surface of the celestial bodies (e.g., the Moon and Mars), as well as georeferenced datasets regarding different thematic contexts such as geology, meteorology, and climate, similar to the SDI concept on Earth.

In this regard, van Gasselt & Nass (2011) developed a data model and entity relationship diagram for storing spatial datasets in the geodatabase in the sense of planetary mapping. This study focused on geology-related datasets in terms of the conceptualization of the data model. Later, establishing a PSDI that focuses on the storing and sharing of planetary-based spatial datasets in an interoperable manner was proposed by different scholars (J. Laura et al., 2017; Nevistić & Bačić, 2022). These studies especially underlined that such infrastructures are highly important on Earth and there is a need for significant consideration for applying similar context on celestial body-related datasets. Regarding this, a case study that considers the SDI for Europa was also presented (J. R. Laura et al., 2018). In this study, it is shown that dispersed information that is generally conveyed between researchers can be rendered as more useful and discoverable through exemplary PSDI. Nass et al. (2021) proposed a conceptualized research-data lifecycle by considering the open map repository in the sense of improving the reusability of planetary spatial research data. The authors emphasized the importance of communication and participation as well as coordination efforts for efficient improvement regarding such a lifecycle, especially for planetary map products. J. R. Laura & Beyer (2021) also provided a catalog of foundational planetary data by covering the different categories, such as resolution and coverage. They underlined that the datasets should be available more easily with clear descriptions, and identifying such data is fundamental for PSDI.

The significance of interoperability was underlined by the scholars since it is one of the vital factors that promote the reusability of planetary spatial datasets and improve analysis capabilities where these datasets are used (Hare, 2023; Hare et al., 2018; Naß et al., 2019).

In terms of data interoperability, standardization of projected coordinate systems for planetary spatial datasets was also highlighted because this issue is notably fundamental for efficient usability of such datasets (H. Hargitai et al., 2019). Scholars also discussed the semantics and metadata regarding planetary mapping by noting the importance of standards and interoperability (van Gasselt & Nass, 2023; van Gasselt & Naß, 2024). They elaborated the relationship between systematic, thematic, and reference mapping domains by considering planetary datasets. A geoportal for planetary geodatasets, including 3D visualizations and GIS-based tools that enable sharing, reaching, and analysing the planetary spatial datasets for different celestial bodies, such as Mars, was also developed to improve the data foundation that is useful for planning other scientific missions (Hare, 2023; Karachevtseva et al., 2018; Minin et al., 2019).

In the sense of spatial datasets, some studies focused on establishing crater databases, for example, identifying craters ≥ 0.4 km on the Moon by using a deep learning model (La Grassa et al., 2025). In addition, there is a growing interest in creating 3D models of celestial bodies. For example, Lu et al. (2021) aimed to provide a 3D model of the Moon including crater information by using the datasets that were obtained from China's Chang'e project. Related to this issue, it can be mentioned that 3D datasets in 3D Tiles[5] format covering both the Moon[6] and Mars[7] surfaces were shared by Cesium. Scholars also focused on developing digital twin applications that benefit from 3D models. In this sense, Bingham et al. (2023) developed a DLES Unreal Simulation Tool (DUST) that has different simulation capabilities such as creating shadow maps and enabling real-time ray tracing to help the exploration of the lunar south pole in a virtual environment. Recently, the authors proposed the use of the IEEE 2874-2025 Spatial Web standard in order to overcome the heterogeneity between different digital twins that are utilized for the lunar surface (Barrio et al., 2026; Carrillo et al., 2025).

### 2.2 Related 3D Spatial Data Standards

CityGML[8], as an Open Geospatial Consortium (OGC) standard, provides the data schema for creating the 3D models of physical objects and logical spaces within the built environment. It has different modules such as Building and Construction in order to enable to creating 3D models with necessary semantics and feature relationships that might be useful for analyses regarding planning and managing the built environment (Kutzner et al., 2020). CityGML Application Domain Extension (ADE) is the mechanism to promote the enhancement of the core data schema of the standard such that it encompasses the feature types and attributes for specific application fields, such as building energy modeling. CityJSON[9] is also proposed as a JSON-based encoding of the core schema of CityGML. The increased use of JSON structure within modern software development, rather than the highly hierarchical and complex nature of Geography Markup Language (GML), is one of the reasons for developing CityJSON (Ledoux et al., 2019). By using the simplified structure within the created datasets, another reason is the possibility of obtaining a smaller file size compared to the CityGML datasets. CityJSON also has an extension mechanism similar to the CityGML ADE concept, yet with a more effortless approach. In other words, the CityJSON datasets that are created in line with the proposed extension can be readily imported and used within different software/tools without the need for an extension file, as CityGML datasets require. In this sense, a number of CityJSON extensions were proposed for different contexts, such as 3D spatial planning (Guler, 2023; Guler et al., 2026), energy demand modeling (Tufan et al., 2022), historical city modeling (Vaienti et al., 2022), integrating human perception (Lei et al., 2024), and design checks (J. Wu, 2021).

---

[5] https://www.ogc.org/standards/3dtiles/

[6] https://cesium.com/platform/cesium-ion/content/cesium-moon/

[7] https://cesium.com/platform/cesium-ion/content/cesium-mars/

[8] https://www.ogc.org/standards/citygml/

In addition to these standards that mainly focused on the modeling of the built environment, Industry Foundation Classes (IFC)[10], as both an open and ISO standard (16739-1:2024[11]), provides a standardization of digital, 3D datasets of the buildings primarily in the context of the Building Information Modeling (BIM) approach. Also, glTF[12] is one of the widely used data formats that allows for delivering the 3D models/assets. Furthermore, 3D Tiles[5] is an OGC community standard that is developed for streaming 3D spatial datasets over web interfaces. More recently, the JSON FG[13] standard that defines GeoJSON extensions for different standardization capabilities, including representing 3D models, has been announced by OGC. It should be emphasized that 3D spatial data standards are highly important for this study since the developed conceptual model that aims to enable the creation of 3D models that cover physical objects and logical spaces can be implemented through these standards.

[9] https://www.cityjson.org/

[10] https://ifc43-docs.standards.buildingsmart.org/

[11] https://www.iso.org/standard/84123.html

[12] https://www.khronos.org/gltf/

[13] https://www.ogc.org/standards/json-fg/

Apart from these data standards, it can be mentioned that the Land Administration Domain Model (LADM), as an ISO standard (19152[14]), that provides a conceptual model for land registry and cadastre-related activities on Earth. With its recently published second edition, this standard covers a variety of land administration counterparts such as land registry, land valuation, and land development by considering 3D delineations of different types of features such as legal spaces representing a parcel, utility network, and building unit (Kara et al., 2024).

### *2.3 Contributions of This Study*

Reviewing the literature shows that there is no solid consensus on property rights regarding space resources and how these resources can be utilized and exploited. The literature also underlines that international agreements and legal frameworks are essential to promote the peaceful management of the activities on celestial bodies from different contexts, such as space heritage. In addition, the need for ensuring interoperability for planetary spatial datasets is evident according to previous studies.

The initial preliminary conceptualization regarding the 3D geoinformation-based modeling of surface objects on the Moon is provided by the study of Guler (2025). However, in that study, only a few surface objects from a limited context were included without considering the related logical spaces. In other words, the mentioned study provided only two feature types that correspond to space objects and restrictions without detailing attributes and relationships. Accordingly, it presented limited delineations, namely 3D models of a small number of craters and selected areas that might contain water, by lacking elaborated descriptions on modeling, transforming, and validating the datasets. For this reason, the content of that study was narrowed a small variety of application/scientific fields, as initially presenting the foundational idea.

Therefore, a research gap has been identified on how to enable strong technical implementation regarding celestial body-related geodatasets that provide the fundamental data sources for the aforementioned issues. In this sense, the present study contributes to existing literature from different points. First, it provides a conceptual model that contains the feature types and their relationships by considering the planning and management of both underground and aboveground surfaces of the celestial bodies. This conceptual model not only covers the physical representation of different surface/subsurface objects, such as craters, but also logical spaces that can be associated with them for different purposes, such as restricted zones for heritage sites. Second, this study provides a notable foundation for how to create 3D standardized geodatasets regarding these surface/subsurface objects and logical spaces through implementing this conceptual model as a CityJSON extension. These datasets that have valuable semantics are of great importance in terms of interoperability for planetary spatial data. Third, the current research paves the way for possible usability of these datasets for space-related activities such as mission planning, since it presents demonstrations regarding the developed standard extension by means of exemplary cases on two celestial bodies, namely the Moon and Mars. Overall, it is expected that the mentioned contributions provide significant resources for designing frameworks for activities regarding the surfaces of the celestial bodies. It is worth noting that the present work uses the above-mentioned study (Guler, 2025) as a foundation and improves the idea significantly through developing a comprehensive standard extension and providing complete details on data modeling, transformation, and visualization.

[14] https://www.iso.org/standard/81263.html

## 3 Methodological Workflow

Figure 1 shows the methodological flow of this study. The first step of the methodology covers utilizing the data requirement analysis to shape the content of the conceptual model for surface/subsurface objects and related logical spaces on celestial bodies. For this step, mentioned issues in the literature, such as space heritage, are considered fundamentally. Developing a conceptual model is the foundational step for creating a usable CityJSON extension. Therefore, required feature types, their attributes, and relationships between these feature types are determined by considering the CityJSON specifications (Ledoux & Dukai, 2023) to better implement the developed conceptual model through standardized 3D geodatasets. Unified Modeling Language (UML)-based approach is followed for designing the conceptual model. In the next step, the CityJSON extension file is created based on the content of the developed conceptual model. This extension file is then validated in terms of its syntax. The details of the mentioned processes are given in Section 4. After obtaining the validated extension file, different cases, such as protecting scientifically important areas, that show the practicability of the proposed model are selected by taking the literature into account. The diverse cases from different domains such as planning, mining, and heritage are evaluated to more completely demonstrate the content of the developed model. Next, 2D spatial datasets representing the selected cases are organized semantically and geometrically. These datasets are then transformed into 3D CityJSON datasets by using FME[15] software. The created CityJSON datasets are also validated against the developed extension file and geometrically through open-source tools. The transformation and validations are elaborated in Section 5. The workflow is completed by providing the 3D visualizations of created and validated CityJSON geodatasets within a variety of tools, such as QGIS.

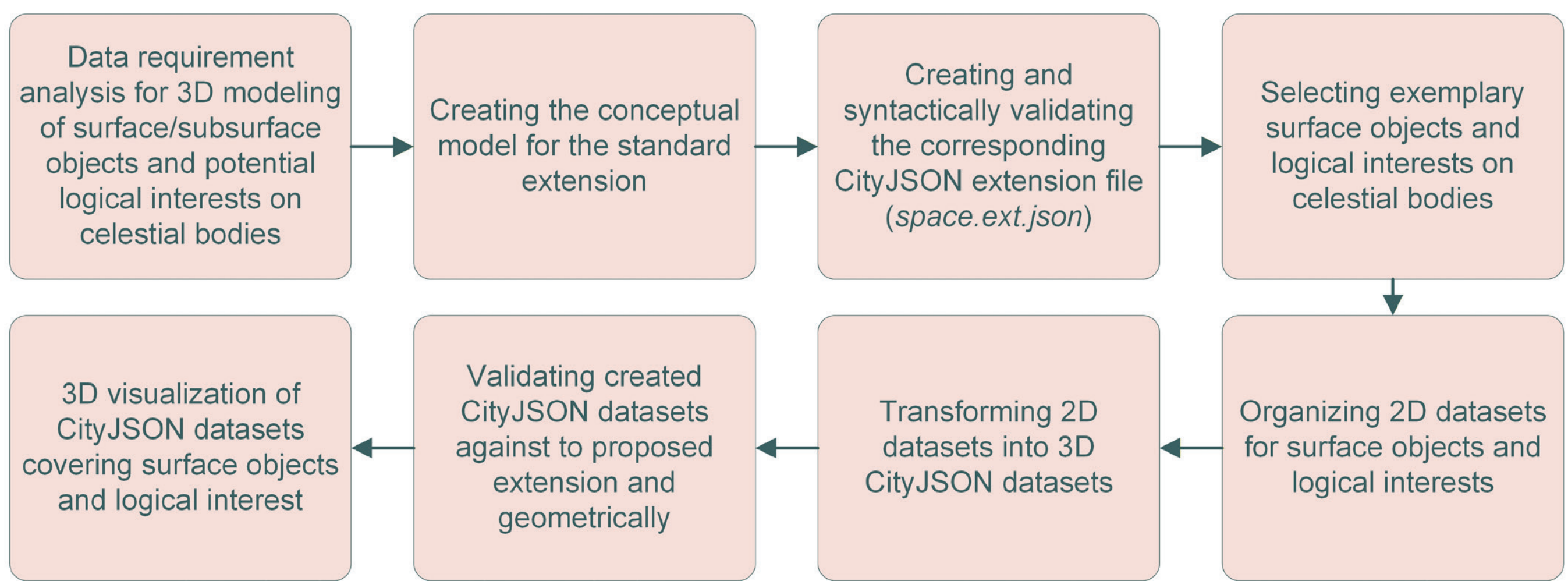

Figure 1. The workflow of this study.

## 4 Developing Conceptual Model and Standard Extension

The first step for creating a standard extension is to develop the conceptual model that covers the required feature types, their attributes, and relationships between these feature types. The defined rules within the data specifications of the CityJSON should be followed during development of the extensions for the core schema of the standard. Accordingly, these specifications should be considered when creating the conceptual model that represents the developed extension. They cover different types of modeling instructions. For example, the names of all newly defined city objects should start with a “+” sign (e.g., +*SpaceSurfaceObject*), and it is not allowed to extend the existing city objects with new city objects as a child. In other words, a new parent city object, such as +*SpaceBuilding,* should be defined within the developed extension in order to add a new child city object, such as +*SpaceBuildingUnit*. The mentioned rules are followed to develop the conceptual model in this study, as can be seen in Figure 2.

---

[15] https://fme.safe.com/

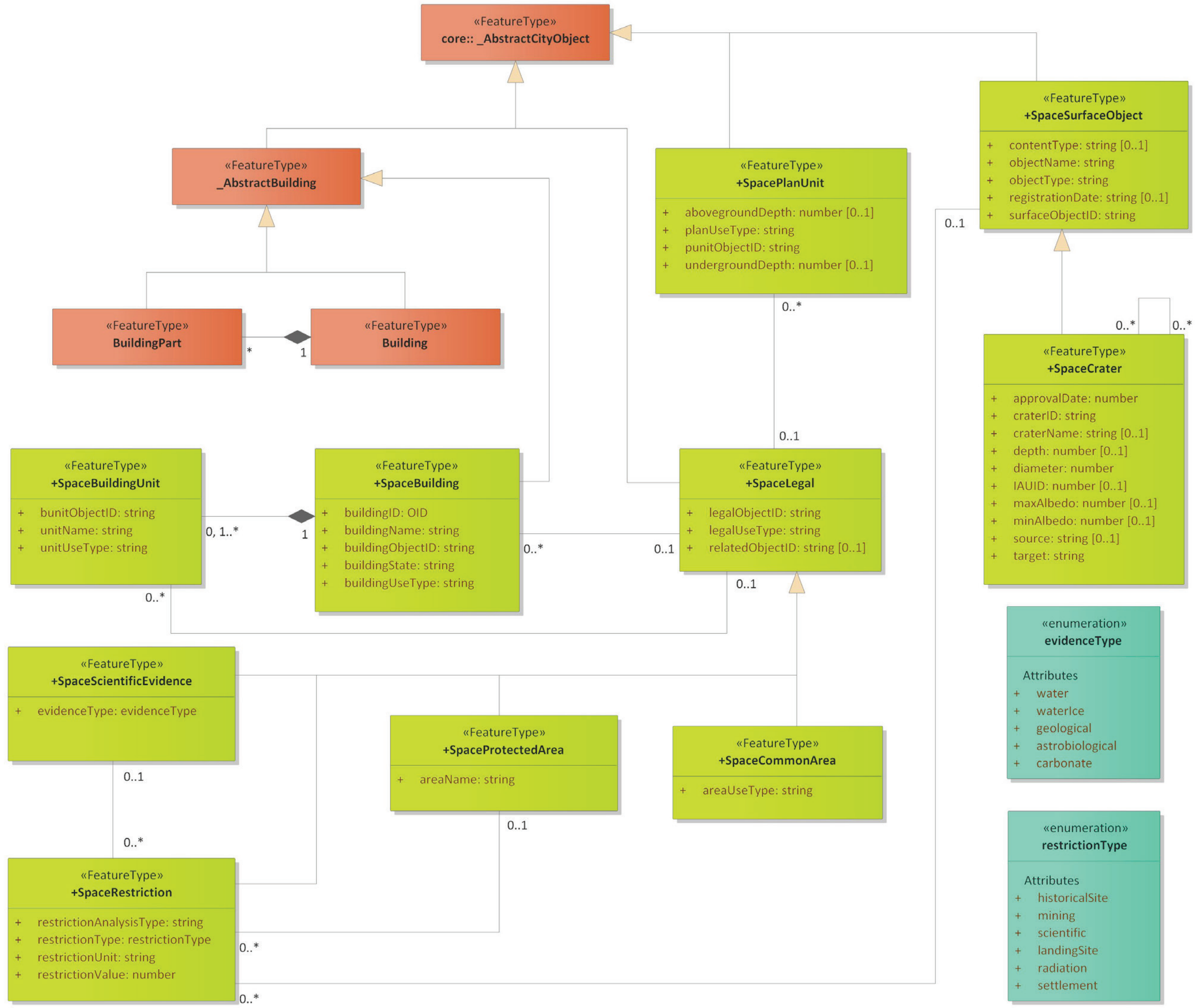

Figure 2. The conceptual model of the developed standard extension that enables 3D modeling of different types of surface/subsurface objects and related logical spaces on celestial bodies.

Data requirement analysis is conducted to identify the essential feature types and attributes that allow for the 3D representation of various space-related surface/subsurface objects and their related logical interests that support both decisions and management of procedures, such as creating a database that includes the historical sites on celestial bodies. Identifying the object types is carried out by exploiting the related studies mentioned in Section 2.1, such as space mining and space heritage. +*SpaceSurfaceObject* is defined to model the different kinds of objects on the surface of celestial bodies, such as scientifically important areas, mountains, faults, and volcanoes. An example of a significant object type that may be described using this kind of feature type is Irregular patches (IP) (H. I. Hargitai & Brož, 2025), which are studied to gather information on the evolution of the Moon. As presented in Figure 2, this object type is a top-level object, and it includes generic attributes such as *objectName*, *objectType*, and *registrationDate* to enable modeling of objects that are needed for different scientific purposes. One of the most noteworthy surface object types is the crater (Yue et al., 2023). For this reason, a specific object type, namely *+SpaceCrater,* is defined in the extension as a subclass of *+SpaceSurfaceObject*. It has more specific attributes compared to the attributes of *+SpaceSurfaceObject* to model the craters in a comprehensive manner.

The content of the crater databases (Robbins, 2019; Wang et al., 2021) and attributes within the Gazetteer of Planetary Nomenclature[16], which is the official naming database for craters, is considered when selecting the attributes of this object type. For example, *+SpaceCrater* has attributes such as *craterID*, *craterName*, and *diameter* to represent the craters in the database in a reusable manner. As illustrated in Figure 2, the existence of some attributes of the object is defined as optional, for example, *depth*. The main reason for this is that some information regarding craters is not commonly available, such as their depths or albedo values.

Another defined feature type is *+SpacePlanUnit* that can be used to model planned land use types of parts of the surface, such as mining. Modeling the locations where the mining activities might be conducted can be given as an example for the possible use of this feature type. These land use types are depicted with the *planUseType* attribute. In addition to this attribute, the *+SpacePlanUnit* feature type might have information regarding the depth information for both above and below the surface. Depth information that represents the planned mining activities can be given as an example in this sense. It also has a *punitObjectID* attribute to identify the *+SpacePlanUnit* instances. *+SpaceBuilding* and *+SpaceBuildingUnit* are the feature types that can be utilized to model buildings and single building units that they cover. One example of such a building can be a lunar settlement facility that might be planned to build in lunar south pole. *+SpaceBuilding* can represent the 3D model of the whole building. Furthermore, building units that are designated for specific purposes, such as scientific experiment unit can be delineated through the *+SpaceBuildingUnit* feature type. As can be seen from Figure 2, these two feature types have different attributes to store the different specifications regarding instances. For example, *+SpaceBuilding* has a *buildingState* attribute that can depict the state of the lifecycle of the building, such as starting, continuing, and finishing the construction. While *buildingID* can be used to depict a unique identification number of a specific building, *buildingObjectID* can express the code of a building that is defined by international organizations, similar to a defined crater ID. As aforementioned, *+SpaceBuildingUnit* instances can be used to model building units with different purposes through defining the *unitUseType* attribute.

[16] https://planetarynames.wr.usgs.gov/

As shown in Figure 2, different relationship types are defined between the feature types. For example, there is a composition relationship between *+SpaceBuilding* and *+SpaceBuildingUnit*. On one hand, a *+SpaceBuildingUnit* instance should have a relationship with only one *+SpaceBuilding* instance. On the other hand, a *+SpaceBuilding* might have a relationship with none, one, or more *+SpaceBuildingUnit* instances. This is because a *+SpaceBuildingUnit* instance cannot exist without a *+SpaceBuilding* instance.

*+SpaceLegal* is the main feature type for 3D modeling of legal/logical spaces regarding the surface/subsurface objects and other logical interests, such as buffer zones for preservation of historical landing sites (e.g., Apollo missions) in the context of space heritage. This feature type represents the volumetric spaces that can be defined by both logical and physical boundaries. As seen in Figure 2, *+SpaceLegal* has various subclasses to thoroughly delineate the different types of logical interests. Among these subclasses, *+SpaceScientificEvidence* can be used to model locations that might have scientific interests. Permanently shadowed regions (PSR) where the scientific evidence regarding water-ice might exist can be given as an example. The different evidence types can be represented through the *evidenceType* attribute, in which their possible content is defined, such as *waterIce*, *geological*, and *astrobiological*. *+SpaceProtectedArea* is another subclass that can be utilized to represent the locations that are decided to be protected, such as locations of previous lunar missions. This feature type has one attribute, namely *areaName,* to store the name of the modeled place. Another subclass is *+SpaceCommonArea,* which can be used to delineate the places where the parties have permission to utilize jointly. Areas where different scientific exploration works can be conducted in parallel are examples of this feature type. It has an *areaUseType* attribute to define the specific use type information for the modeled location.

Another subclass is *+SpaceRestriction,* which can be used to model different kinds of restrictions regarding various logical interests. For this reason, this feature type has relationships with other feature types, that is, *+SpaceScientificEvidence,* *+SpaceProtectedArea,* and *+SpaceSurfaceObject*. There exist 0..* and 0..1 relationships between the mentioned feature types. This means that a number of restrictions can be defined and modeled for different types of areas, such as scientific evidence locations and protected areas. *+SpaceRestriction* has several attributes to provide details regarding the defined restrictions. For instance, *restrictionAnalysisType* expresses the type of analysis conducted, such as a 2D buffer. Additionally, the value and unit information regarding applied analysis can be stored by using *restrictionUnit* and *restrictionValue* attributes, respectively. The different types of restrictions can be selected through the *restrictionType* attribute, such as *historicalSite*, *mining*, *scientific*, and *settlement*. Restrictions that are modeled for possible lunar settlement, PSRs, and IPs can be given as an example for possible usability of this feature type.

Figure 2 also illustrates the relationships between *+SpaceLegal* and other feature types namely *+SpaceBuildingUnit,* *+SpaceBuilding* and *+SpacePlanUnit.* These relationships are defined within the conceptual model to enable 3D modeling of logical interests pertaining to different types of features. For example, a 3D representation of logical space regarding an area that is defined as a mining location through *+SpacePlanUnit* can be modeled by using *+SpaceLegal* feature type. Therefore, it has several attributes, such as *relatedObjectID,* in order to provide the connection between the feature instance and the related logical space. This attribute is also inherited in other subclasses of *+SpaceLegal,* such as *+SpaceRestriction,* for a similar objective.

Noteworthy to mention is that this study includes the 3D modeling of logical spaces that represent the possible/suggested mining areas on celestial bodies. It doesn't focus on the mining extractions. This is because the concept that is followed in this study is similar to land administration studies on Earth, whose main focus is 3D modeling of logical spaces that represent the right of use through standardized datasets.

The rules that are defined for developing the CityJSON extension are followed when creating the corresponding extension file for the conceptual model in Figure 2. Accordingly, the relationships within the conceptual model are defined within the extension file. For instance, *+SpaceBuilding* is defined as a subclass of *_AbstractBuilding* feature type in the core schema of CityJSON. *+SpaceLegal,* *+SpacePlanUnit,* and *+SpaceSurfaceObject* are defined as subclasses of *_AbstractCityObject,* which is the core object type within the standard's schema. Such a relationship can be seen in Figure 3a.

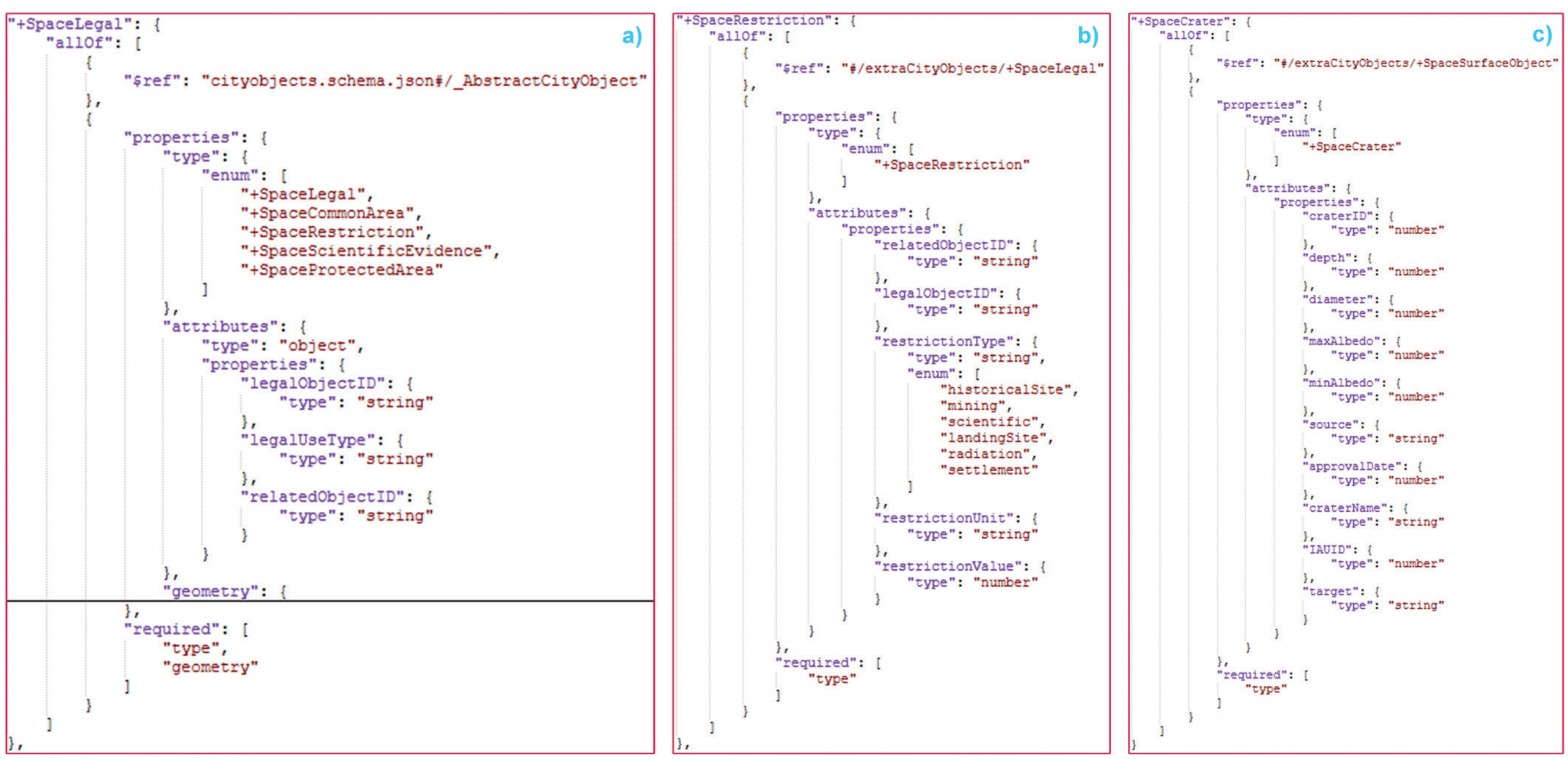

Figure 3. The selected parts of the created CityJSON extension file.

*+SpaceScientificEvidence, +SpaceRestriction, +SpaceProtectedArea,* and *+SpaceCommonArea* are defined as subclasses of *+SpaceLegal.* Figure 3a also shows this relationship since the property types of these feature types are defined for *+SpaceLegal.* Figure 3b illustrates the part of the created CityJSON extension belonging to *+SpaceRestriction.* As depicted in this figure, *+SpaceLegal* is referenced as an upper-level class for *+SpaceRestriction.* Figure 3c indicates that *+SpaceCrater* is modeled as a subclass of *+SpaceSurfaceObject*, as abovementioned before. These three figures also present that the attributes of the feature types are included in the developed extension file in line with the content of the conceptual model in Figure 2. Whereas Figure 3a shows that geometry is defined as a required property for *+SpaceLegal,* Figure 3a and Figure 3b illustrate that geometry is not required for *+SpaceRestriction* and *+SpaceCrater.* This is because these two feature types are modeled as subclasses and hence the geometry specifications can inherit from their upper-level classes, as is the case for *+SpaceLegal* and *+SpaceRestriction*. The created extension file is validated in terms of JSON syntax and then shared in the open repository[17] of this study to make it possible to validate the CityJSON datasets that are created according to the developed extension. The details for this validation are given in Section 5.

[17] https://github.com/geospatialstudies/space

# 5 Demonstration

This section presents the demonstration of the usefulness of the proposed standard extension through creating CityJSON datasets that cover the 3D modeling of exemplary object types and related logical interests.

## *5.1 Description of Exemplary Cases*

Table 1 lists the types of exemplary surface/subsurface objects, their target celestial bodies, their object types in the proposed extension, and details on how they are modeled.

Table 1. Modeling details for the selected surface/subsurface objects

| **Target** | **Type** | **Object Type in Extension** | **Modeling Detail of the Demonstration** |
|---|---|---|---|
| Moon | Craters | +*SpaceCrater* | Incorporating with DEM |
| Moon | Planned mining areas | +*SpacePlanUnit* | - |
| Moon | Exemplary building | +*SpaceBuilding* | - |
| Moon | Exemplary building units | +*SpaceBuildingUnit* | - |
| Moon | Permanently shadowed regions (PSR) | +*SpaceScientificEvidence* | Underground and aboveground extrusion (25 m) |
| Moon | Irregular patches (IP) | +*SpaceSurfaceObject* | Underground and aboveground extrusion (1 m) |
| Moon | Historical landing sites | +*SpaceProtectedArea* | Underground and aboveground extrusion (1 m), Buffer (5 m) |
| Moon | Possible lunar settlement | +*SpacePlanUnit* | Underground and aboveground extrusion (50 m) |
| Mars | Home Plate in the Columbia Hills of Gusev Crater | +*SpaceScientificEvidence* | Underground and aboveground extrusion (2.5 m), Buffer (1 m) |
| Mars | Crater | +*SpaceCrater* | Incorporating with DEM |

As shown in Table 1, there is a variety of surface/subsurface object types in which the existence of their 3D models promotes sustainable space exploration in the sense of data-driven planning and management. To present the applicability of the proposed extension within different study areas, demonstrations for both Moon and Mars are included in this study. Most of the selected surface/subsurface types are located in the lunar south pole region, which is one of the most significant places on the Moon (Reach et al., 2023). This is because it contains highly valuable scientific evidence, and it is aimed at scientific explorations within a great number of space missions. One of the important surface object types is the crater. For this reason, three craters on the lunar south pole, namely Shackleton, Shoemaker, and Tooley, are selected to show how a 3D standardized geodataset representing the craters can be created. Space mining is one of the most discussed topics in the literature. The discussions encompass which locations are suitable for mining activities and how to plan the mining activities that will be conducted (Steffen, 2022). For this reason, planned mining areas where the proposed extension can be beneficial in terms of enabling a spatial dataset are included in this study. In addition, two connected surface/subsurface object types might be of interest for managing the human bases on celestial bodies. Similar to the digital depiction of the buildings and their units (e.g., apartment units) on the built environment on Earth, these bases are covered as buildings and building units on the surface or subsurface of the celestial bodies within the demonstration for the proposed extension.

PSR are another significant surface type that is of great importance for scientific evidence, such as water ice (Lemelin et al., 2021), and thereby creating their 3D models in a standardized manner would be useful for planning exploration studies and protecting these areas. Thus, a selection of PSR within the lunar south pole is included in this research. In addition, IP are another important surface object type since how they are formed can provide vital insight into the Moon's evolution (H. I. Hargitai & Brož, 2025). Accordingly, they are of great scientific interest, and hence the investigation on how 3D geodatasets of them and their possible protection buffers can be created is covered in the current study. Another demonstration consists of historical landing sites in which there exist strong opinions for protecting these sites in the context of space heritage (Walsh, 2012). 3D geodatasets that represent the historical places with their attributes would be useful for both creating the inventory on space heritage and storing these places in an interoperable way. Furthermore, establishing a settlement on celestial bodies is discussed within future space missions. A lunar settlement/base is one of the near-future aims for different countries (Peña-Asensio et al., 2025). For this reason, a possible lunar settlement is selected as an exemplary surface/subsurface object. The aforementioned surface/subsurface objects are targeted on the Moon. In addition, Jezero crater, which is of scientific interest on Mars in terms of biosignatures, is selected for demonstration purposes (Mangold et al., 2021;

Tarnas et al., 2021). Another important area on Mars, namely Home Plate in the Columbia Hills of Gusev Crater, is included because it is of high scientific interest for different aspects, such as astrobiology (Ruff & Farmer, 2016).

### *5.2 Data Modeling and Transformation*

As seen in Figure 1, the research workflow includes organizing 2D spatial datasets for selected surface/subsurface objects in Table 1. Figure 4 shows the map of selected surface/subsurface objects for craters, settlement, PSRs, and mining areas. There are two basemaps in this figure. The first one is the LROC WAC Global Morphologic Map[18] at 100 m resolution (Speyerer et al., 2011). The second one is the South Pole LOLA DEM Mosaic[19] that is created for the selected region in the lunar south pole at high resolution (i.e., 5mpp) to support scientific missions (Barker et al., 2021).

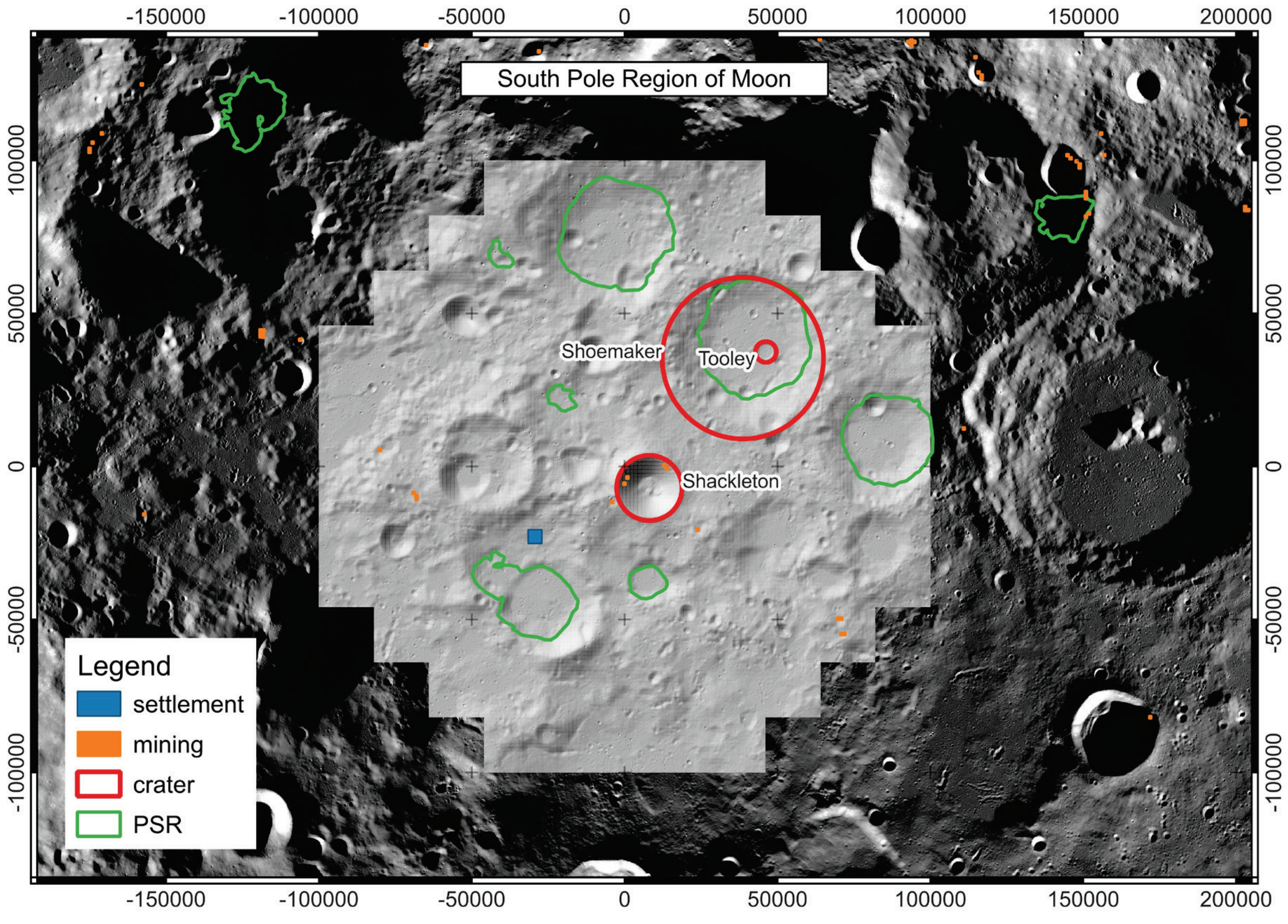

Figure 4. The map that shows selected craters and PSR, and the location of possible human settlement and mining locations.

---

[18] https://data.lroc.im-ldi.com/lroc/view_rdr/WAC_GLOBAL

[19] https://pgda.gsfc.nasa.gov/products/81

A 2D spatial dataset in polygon type that represents the three craters on the Moon is created based on the open crater database[20] that is shared by Wang et al. (2021). The diameters of these craters are also checked based on the Gazetteer of Planetary Nomenclature. Another dataset[21] that covers the identified suitable mining areas on the lunar south pole as polygons, namely grids in 1 km x 1 km, is shared by Hubbard et al. (2024). They identified suitable mining areas based on different criteria, such as an average maximum summer temperature and slope. Lemelin et al. (2021) identified the 37 PSRs on the lunar south pole as sites of interest, and 11 PSRs among them are selected in a way that they can meet the demands for at least one scientific objective. In the present study, the 2D spatial dataset[22] in polygon type of these 11 PSRs is used for demonstration purposes. Leone et al. (2023) identified a location as one of the suitable candidates for a lunar base on the south pole. The coordinates of this location are shared as 88.76°S–232.00°W and are located on the north-eastern side of the Sverdrup-Henson crater. By using these coordinates, in the current study, a polygon dataset for possible lunar settlement is created as a 4.5 km x 4.5 km grid, as suggested by that paper. As shown in Figure 4, the abovementioned surface/subsurface objects are located within the lunar south pole region. Figure 5 illustrates the selected historical landing sites and IPs, which are located in relatively close regions. The bounding box that covers the selected IPs can be seen on the map on the left of Figure 5. To achieve the data transformation, the projection system is defined as IAU2015:30185 Moon (2015) - Sphere / Ocentric / Albers Equal Area for the spatial dataset that covers the landing sites and IPs. The coordinates of these sites are obtained as latitude and longitude from Wagner et al. (2017). Table 2 lists the selected landing sites and their coordinates in the defined projection system.

H. I. Hargitai & Brož (2025) identified the clusters of IPs on the Moon and shared the polygon dataset covering these IPs in Keyhole Markup Language (KML) format (see supplementary material 1 in the mentioned paper). Within this dataset, the group of IPs named Secchi is selected for demonstration purposes.

---

[20] https://doi.org/10.5281/zenodo.4983248

[21] https://dataverse.harvard.edu/dataset.xhtml?persistentId=doi:10.7910/DVN/Q1Y3OS

[22] https://doi.org/10.5281/zenodo.4646092

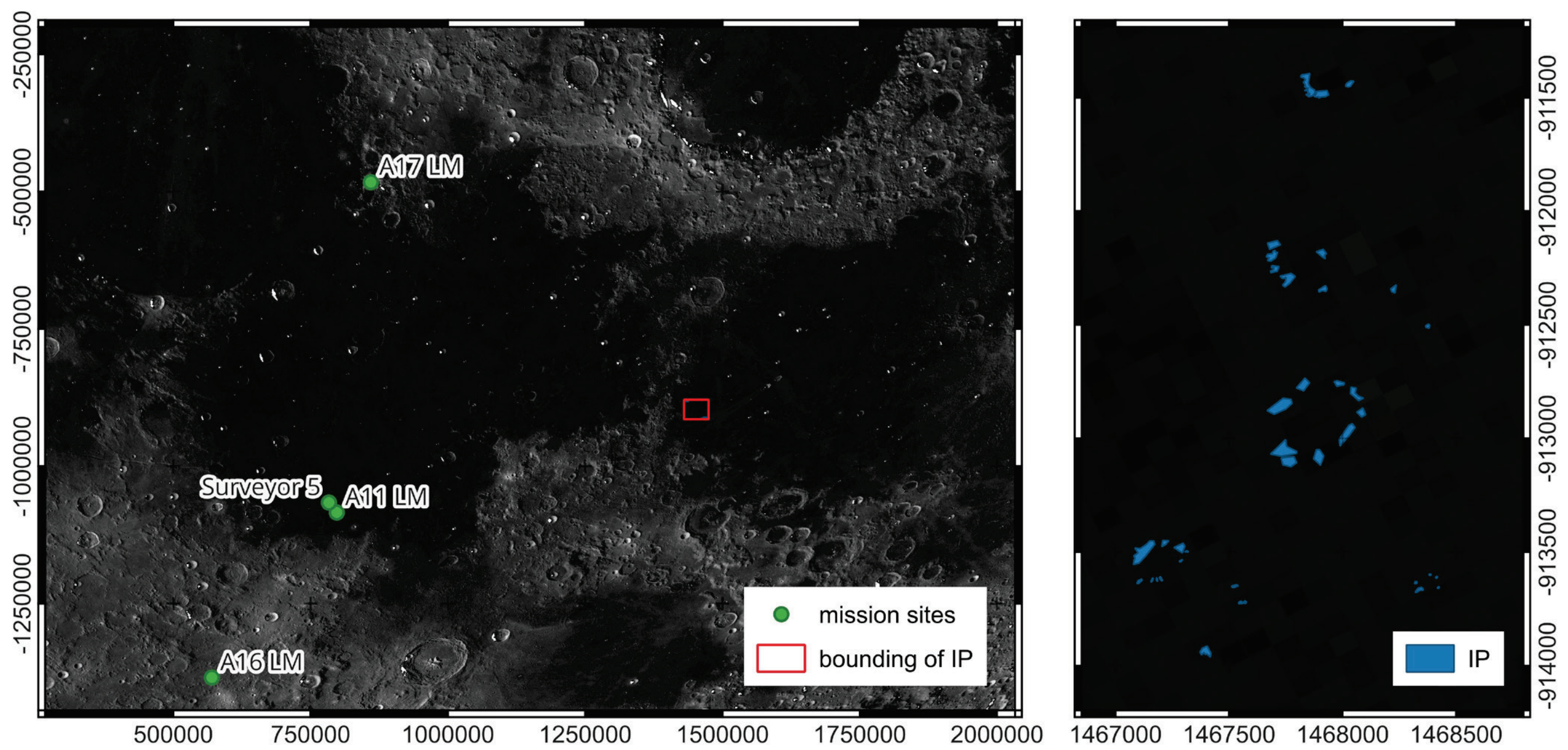

Figure 5.The map that shows the selected mission sites and IPs.

Table 2. The selected historical landing sites and their coordinates (LM: Lunar Module)

| Site Name | Latitude | Longitude | Y | X |
|---|---|---|---|---|
| Apollo 11 LM | 0.67416 | 23.47314 | -1084015.403 | 797715.8357 |
| Apollo17 LM | 20.19106 | 30.77228 | -482883.3008 | 859698.0733 |
| Apollo12 LM | -3.01279 | -23.42192 | -1178748.525 | -819780.9946 |
| Apollo14 LM | -3.64589 | -17.47194 | -1239858.902 | -617302.2113 |
| Apollo15 LM | 26.13239 | 3.63330 | -438959.4181 | 96283.52627 |
| Apollo16 LM | -8.97344 | 15.50105 | -1384487.867 | 570180.2265 |
| Surveyor 1 | -2.47448 | 316.66020 | -923865.0175 | -1473497.873 |
| Surveyor 3 | -3.01623 | -23.41801 | -1178868.968 | -819668.6523 |
| Surveyor 5 | 1.45515 | 23.19426 | -1065991.24 | 783350.0884 |
| Surveyor 6 | 0.47424 | -1.42752 | -1188202.836 | -49090.49337 |
| Surveyor 7 | -40.98117 | -11.51270 | -2052661.123 | -503268.1364 |
| Chang'e 3 | 44.12142 | -19.51174 | 173854.6813 | -396367.0196 |
| Yutu Rover | 44.12085 | -19.51219 | 173838.3572 | -396379.8343 |

Afterward, it is organized as a 2D spatial dataset with the abovementioned projection (CRS 30185). In addition, the IFC dataset, which is an interoperable format within BIM, of a building is used to demonstrate the 3D modeling of a building and building units on celestial bodies. The footprint of this dataset couldn't be shown within the 2D maps (e.g., Figure 4) since the size of its boundaries is significantly small compared to the other surface objects, such as craters.

Figure 6 presents the selected surface objects on Mars, namely Jezero crater and the Home Plate region.

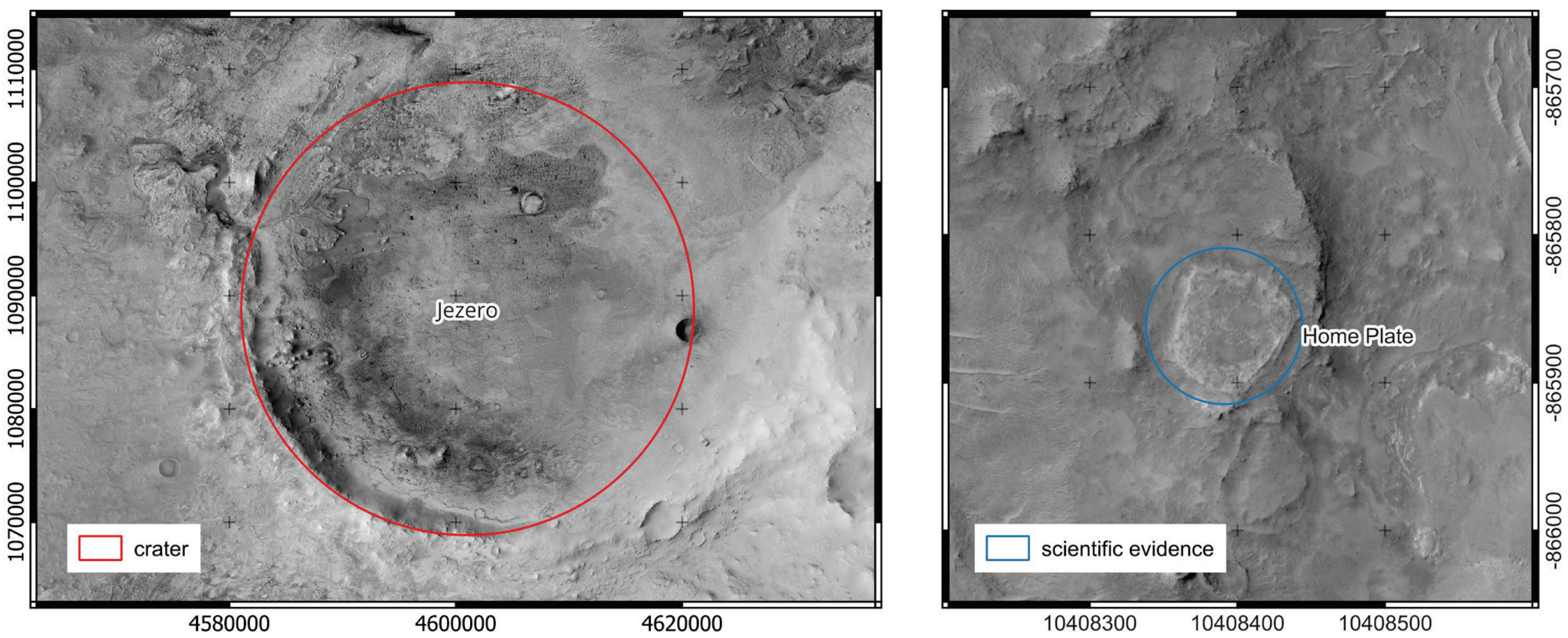

Figure 6. The map that shows the selected surface objects on Mars.

The spatial dataset in polygon type belonging to Jezero crater is created based on the center location and its diameter that are shared within the Gazetteer of Planetary Nomenclature. The polygon dataset that represents the Home Plate is created based on the previous studies (Arvidson et al., 2010; Fletcher et al., 2024). While the map on the left of Figure 6 contains the MurrayLab_CTX_V01_E076_N16_Mosaic[23] as a base map with 5 m resolution (Dickson et al., 2024), the map on the right includes the HIRISE image with the ID of PSP_001513_1655_RED[24], which has 0.25 m resolution. The CRS of the aforementioned 2D spatial datasets is set to 103885 - Mars 2000 Equidistant Cylindrical (sphere) to be able to conduct the data transformation efficiently.

In the next step, as illustrated in Figure 1, transforming 2D spatial datasets that delineate the different surface/subsurface objects into 3D CityJSON datasets is carried out by means of FME software, which is an Extract, Transform, and Load (ETL) tool. This means that it allows for the amending of data on both attributes and geometries. The FME Workbench 2025.1 version is used in this study. This version can write CityJSON 1.0 datasets. As listed in Table 1, different object types are matched with the feature types in the proposed extension. For example, crater objects are converted to 3D models as the *+SpaceCrater* feature type, which is depicted in Figure 2. Regarding this, Figure 7 shows an excerpt of the created workbench that enables to creating *+SpaceCrater* instances within the CityJSON dataset.

[23] https://murray-lab.caltech.edu/CTX/V01/tiles/

[24] https://www.uahirise.org/hiwish/browse

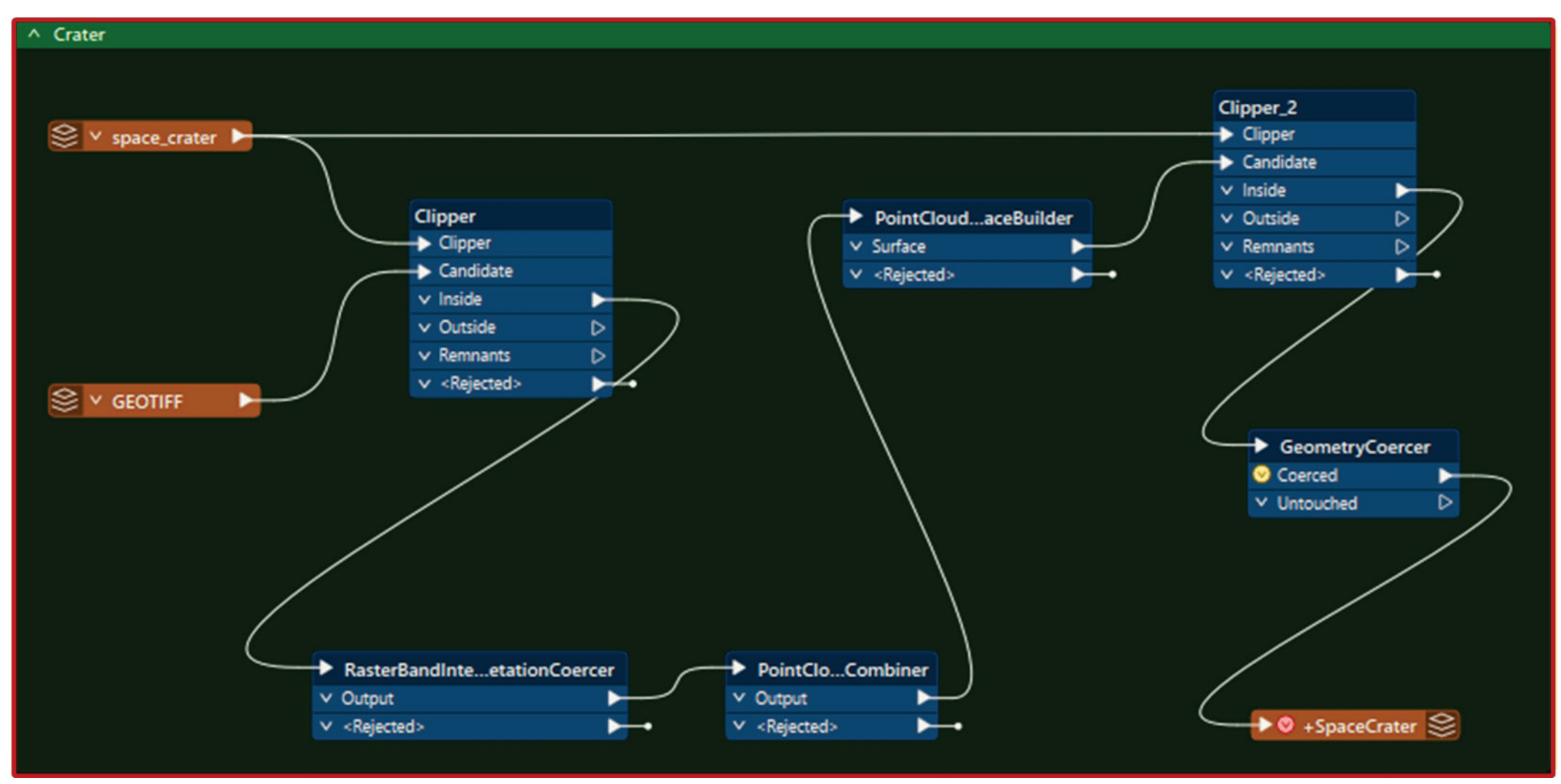

Figure 7. The part of the FME workbench that is used for creating a 3D model of craters.

This part uses a 2D spatial dataset that represents the crater objects and a DEM dataset that covers the boundaries of these craters. For instance, a 2D spatial dataset that includes the polygons of three selected craters is used as input. In this case, the South Pole LOLA DEM Mosaic is aggregated into 100 m spatial resolution and used as a DEM input because its original resolution (e.g., 5 m) requires a large amount of computer memory during the creation of 3D models. As seen in Figure 7, the workbench first clips the input DEM dataset based on the polygons of craters to decrease the computing effort. Then, it transforms clipped DEM datasets into a point cloud dataset, and hence, the surface based on this point cloud dataset is created. In the next step, the created surface is clipped based on the polygons of the crater again, since it covers the bounding box that includes all input craters. At the end, the type of geometry is defined.

This workbench is also used for creating a 3D model of Jezero crater. In this case, the DTM dataset with tile ID of HMC_13E10_da5[25] that is created using the High Resolution Stereo

[25] https://maps.planet.fu-berlin.de/#map=10/4611406.04/1091535.86

Camera (HRSC), is used as input. As indicated in Table 1, mining areas in Figure 4 are modeled as *+SpacePlanUnit* feature type since they represent the planned areas. PSRs are modeled as *+SpaceScientificEvidence* because they express the areas that are of scientific interest. To represent these areas in 3D rather than 2D, the extrusion for underground and aboveground directions is applied, as listed in Table 1. Likewise, the Home Plate region on Mars is modeled as *+SpaceScientificEvidence* with extrusions. Similar to mining areas, a possible lunar settlement location is modeled as *+SpacePlanUnit.* However, the extrusion is applied since the settlement can be used within both above and below of the surface. IPs that represent the surface areas are modeled as *+SpaceSurfaceObject.* Historical landing sites that should be the topic of preservation are modeled using *+SpaceProtectedArea*. For initial protection purposes, both extrusions and a small amount of buffer (i.e., 5 m) are applied during modeling of these areas. The exemplary building and building units are modeled as *+SpaceBuilding* and *+SpaceBuildingUnit* feature types, respectively. The abovementioned feature types are used for creating 3D models of different surface/subsurface objects. In addition to this, the logical spaces related to these surface/subsurface objects are modeled through the feature types in the proposed extension. Table 3 itemizes these feature types and modeling details.

Table 3. Modeling details for related logical spaces of the selected surface/subsurface objects

| **Target** | **Type** | **Object Type in Extension** | **Modeling Detail of the Demonstration** |
|---|---|---|---|
| Moon | Planned mining areas | *+SpaceLegal* | Underground extrusion (500 m) |
| Moon | Exemplary building | *+SpaceLegal* | 3D Buffer (0.001 m) |
| Moon | Exemplary building units | *+SpaceLegal* | 3D Buffer (0.001 m) |
| Moon | Permanently shadowed regions (PSR) | *+SpaceRestriction* | 3D buffer (250 m) |
| Moon | Irregular patches (IP) | *+SpaceRestriction* | 3D buffer (10 m) |
| Moon | Historical landing sites | *+SpaceRestriction* | Underground and aboveground extrusion (25 m), Buffer (75 m) |
| Moon | Possible lunar settlement | *+SpaceRestriction* | 3D buffer (50 m) |
| Mars | Home Plate in the Columbia Hills of Gusev Crater | *+SpaceRestriction* | Underground and aboveground extrusion (25 m), Buffer (250 m) |

For example, the planned mining areas are represented with the *+SpacePlanUnit* feature type; however, there is a need for planning and managing the mining activities by considering the subsurface. Therefore, logical space that might be preserved for these activities can be modeled with the *+SpaceLegal* feature type. As indicated in Table 3, underground extrusion (i.e., 500 m) is applied for creating logical space regarding planned mining activities. The amount of extrusion is selected archetypally for demonstration purposes. As outlined in previous sections, the restriction zones should be applied to different areas on celestial bodies. Moreover, delineating these zones in 3D would be more beneficial. This is because they might be affected by external sources in 3D and also might contain significant scientific evidence on the subsurface. Thus, *+SpaceRestriction* is used for modeling the related logical spaces regarding different surface/subsurface objects by applying 3D analysis, such as a 3D buffer, as shown in Table 3. In particular, a restriction to preserve the historical landing sites is modeled by using underground and aboveground extrusion and buffer analysis as well. Different size of buffers is applied for protecting the areas that are of significant scientific interest, such as PSRs and IPs. In addition, to manage the logical spaces within the building and building units that might be exploited by different parties, these spaces are modeled by means of the *+SpaceLegal* feature type with a 3D buffer analysis. Figure 8 shows the excerpt from the created workbench that converts a 2D spatial dataset representing historical landing sites into *+SpaceProtectedArea* and *+SpaceRestriction* features based on the modeling details that are listed in Table 1 and Table 3.

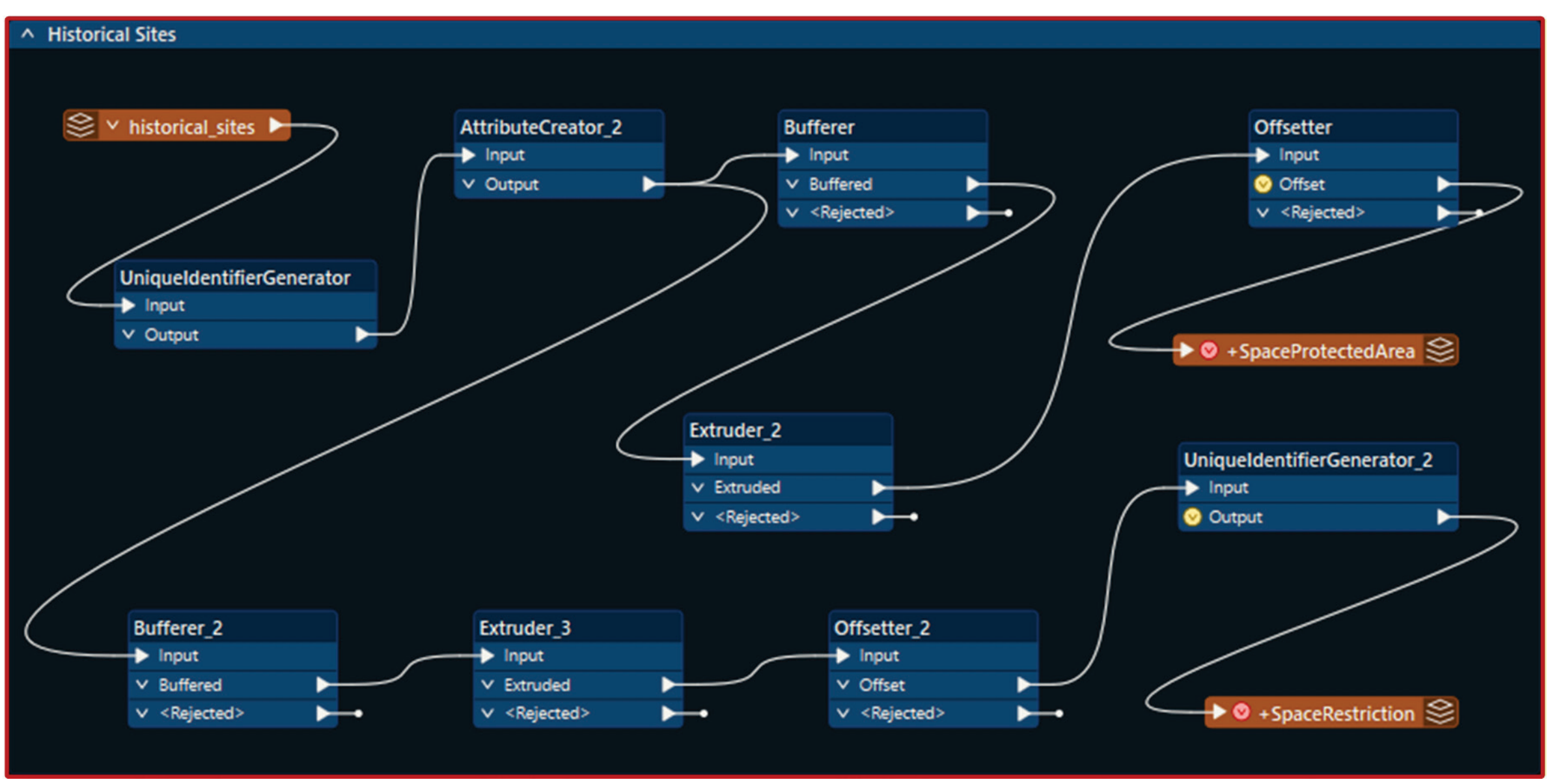

Figure 8. The part of the workbench that creates *+SpaceProtectedArea* and *+SpaceRestriction* features.

It is important to note that the relationship between the surface/subsurface object and related logical space is defined within the proposed conceptual model, as illustrated in Figure 2. To realize this relationship in the created CityJSON dataset, the unique ID of a surface/subsurface object, such as a historical landing site, is added as an attribute titled *relatedObjectID* to the related logical space feature, such as *+SpaceRestriction.* A similar realization between *+SpaceProtectedArea* and *+SpaceRestriction* is included in the workbench in Figure 8. Since some of the exemplary surface/subsurface objects on the Moon are located in highly different regions, and also there are examples from Mars, three different CityJSON datasets are created by means of the developed FME workbenches. These datasets are CityJSON 1.0 because the current version of FME can write only this version of the standard, as mentioned before. However, the standard extension in this study is developed based on CityJSON 2.0. Because of this, these datasets are upgraded to CityJSON 2.0 by using *cjio*[26], which is a command-line interface (CLI) shared by the developers of the standard. Noteworthy to mention is that one of the important steps regarding developing a standard extension is to check whether the created CityJSON datasets are in line with this extension. Accordingly, three CityJSON datasets are validated without error using the *cjval*[27], which is the official validator of the standard. This tool verifies that the CityJSON dataset complies with the core schema, as well as the provided extension that is defined within the CityJSON dataset. This definition is done by adding the openly accessible link of the developed extension file (JSON) in the open repository into the created CityJSON datasets. By doing so, the validator tool can access the content of the proposed extension in terms of feature types, attributes, and relationships so that it checks the compatibility of the CityJSON dataset. In addition, three CityJSON datasets that are shared within the open repository[17] are validated geometrically without any error by means of the *val3dity*[28] open tool (Ledoux, 2018) that enables the validation of 3D primitives against the international standard ISO19107[29]. The reports showing validation results can be seen in the same repository.

---

[26] https://github.com/cityjson/cjio

[27] https://validator.cityjson.org/

[28] https://val3dity.readthedocs.io/2.6.0/

[29] https://www.iso.org/standard/66175.html

### 5.3 *Data Visualization*

Figure 9 presents the 3D visualization of crater objects in QGIS. This visualization is done by the CityJSON Loader plugin (Vitalis, Arroyo Ohori, et al., 2020). As depicted in Figure 9, craters are modeled as instances of the +*SpaceCrater* feature type.

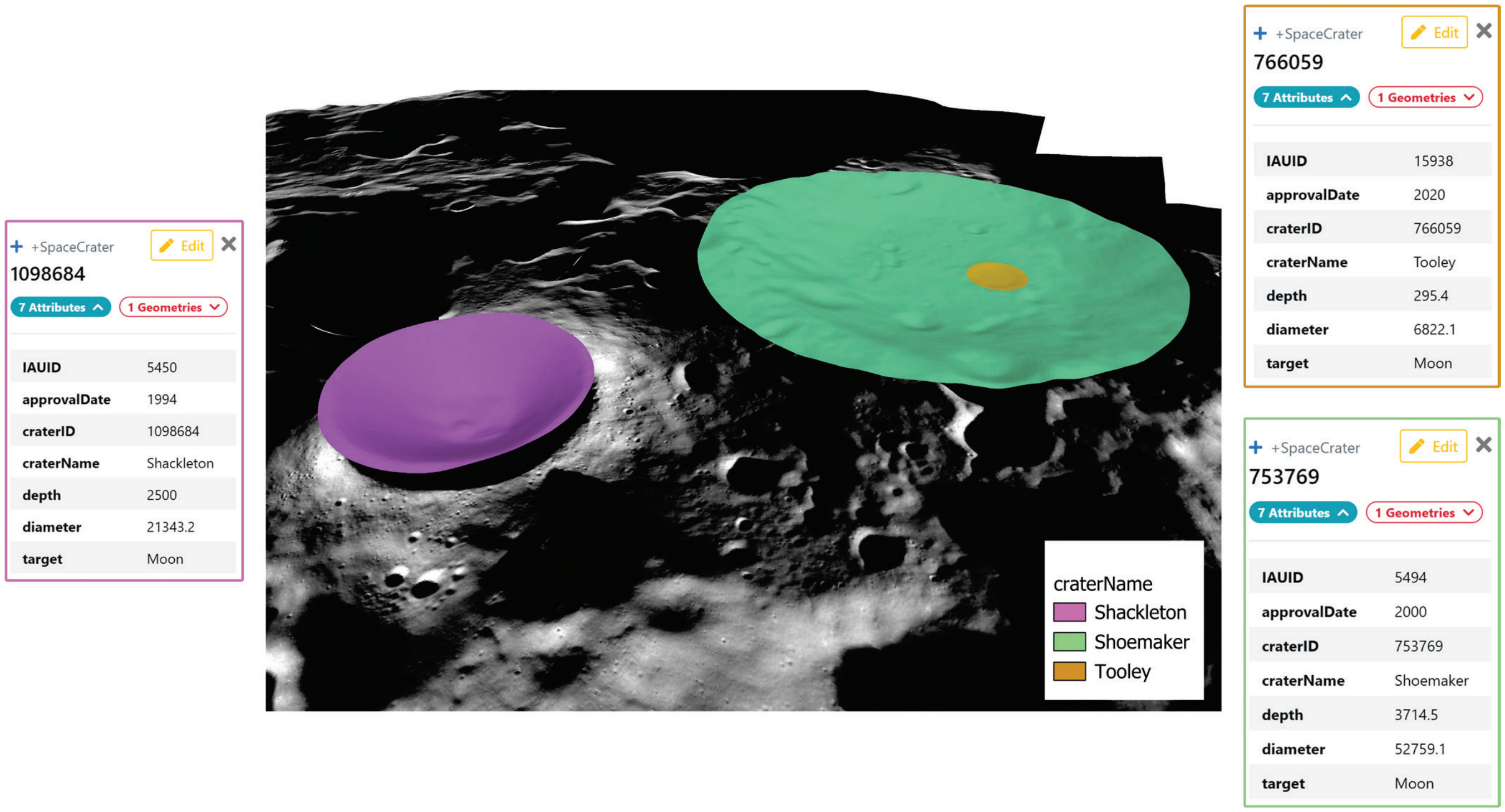

Figure 9. 3D visualization of created +*SpaceCrater* objects. The border colors of the boxes are matched with the color of the objects.

This figure also illustrates the attributes of these +*SpaceCrater* objects. It can be noted that the names of the attributes are matched with the content of the conceptual model of the proposed extension in Figure 2. The values of the *craterID* and *depth* attributes are obtained from the crater database. Other attribute values, such as *IAUID,* are populated based on the Gazetteer of Planetary Nomenclature. One of the reasons to select these craters as an example is to show the usefulness of the proposed approach for modeling two craters that overlap. As indicated in Figure 9, Shoemaker and Tooley craters represent the mentioned situation, but they are able to be modelled as two different features in a standardized way. Figure 10a and Figure 10b show the 3D visualization (above and below ground views) of +*SpacePlanUnit* and +*SpaceLegal* instances that represent the selected mining areas. Two indicated +*SpacePlanUnit* instances in Figure 10a have the same attributes in a way to comply with the proposed extension. Their values for the *planUseType* attribute, namely *mining,* express that these objects represent the planned mining areas.

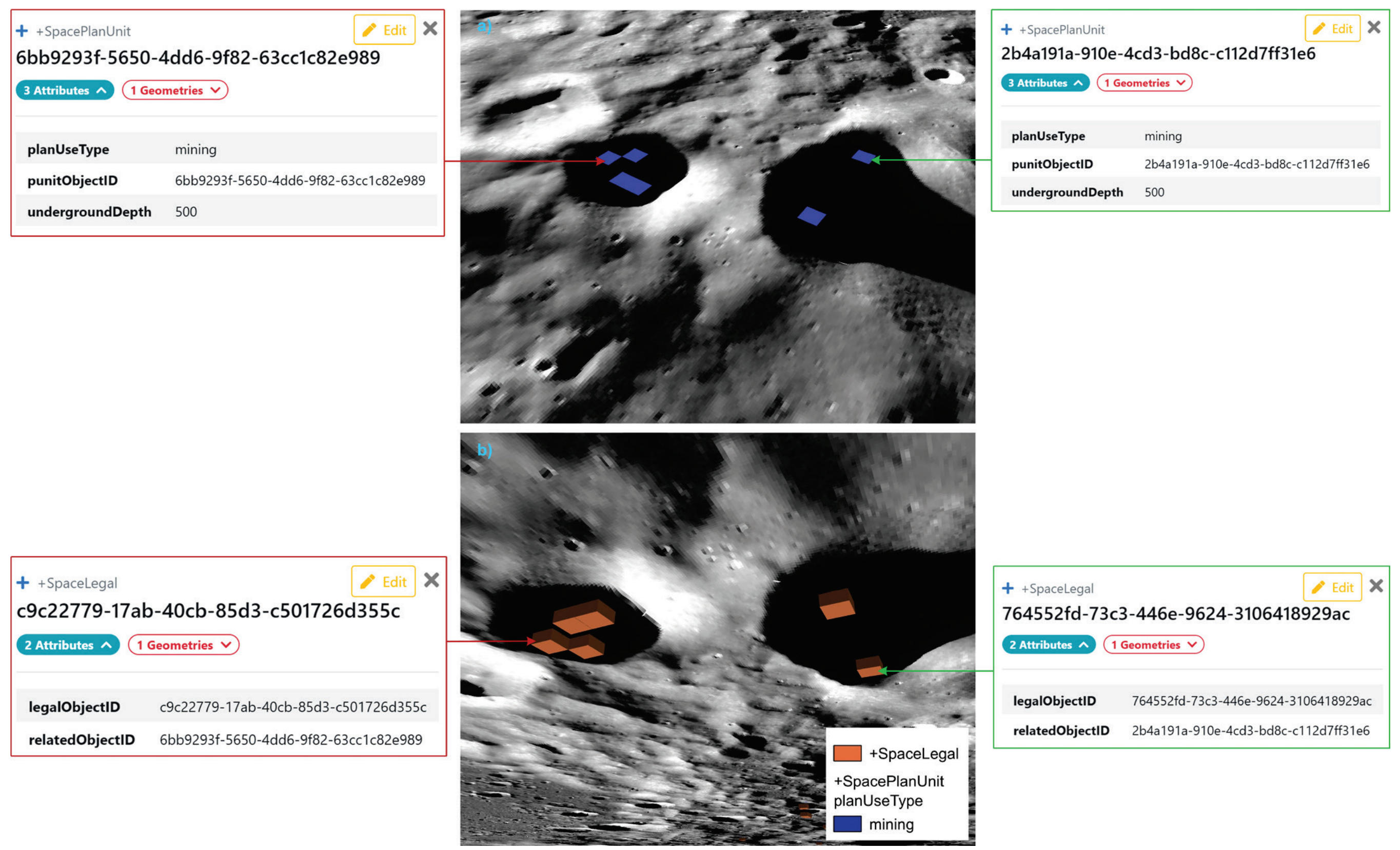

Figure 10. 3D visualization of created +*SpacePlanUnit* and +*SpaceLegal* objects regarding mining areas from surface (a) and subsurface (b) views.

The planned depth for mining activities in these areas is also indicated through the *undergroundDepth* attribute. As mentioned before, logical spaces related to mining activities can be modeled in 3D by means of the proposed extension. Regarding this, Figure 10b illustrates the 3D models belonging to +*SpaceLegal* instances. These instances are connected to +*SpacePlanUnit* instances in Figure 10a. For example, there exists a +*SpacePlanUnit* instance with ID “6bb9293f-5650-4dd6-9f82-63cc1c82e989” in Figure 10a, and Figure 10b contains the related +*SpaceLegal* instance that has the same value for the *relatedObjectID* attribute as the ID of this +*SpacePlanUnit* instance. The same connection can be seen in other indicated +*SpacePlanUnit* and +*SpaceLegal* instances in Figure 10.

Figure 11 also shows the 3D visualization of +*SpacePlanUnit* instance; however, this instance expresses the settlement with the value of *planUseType* attribute.

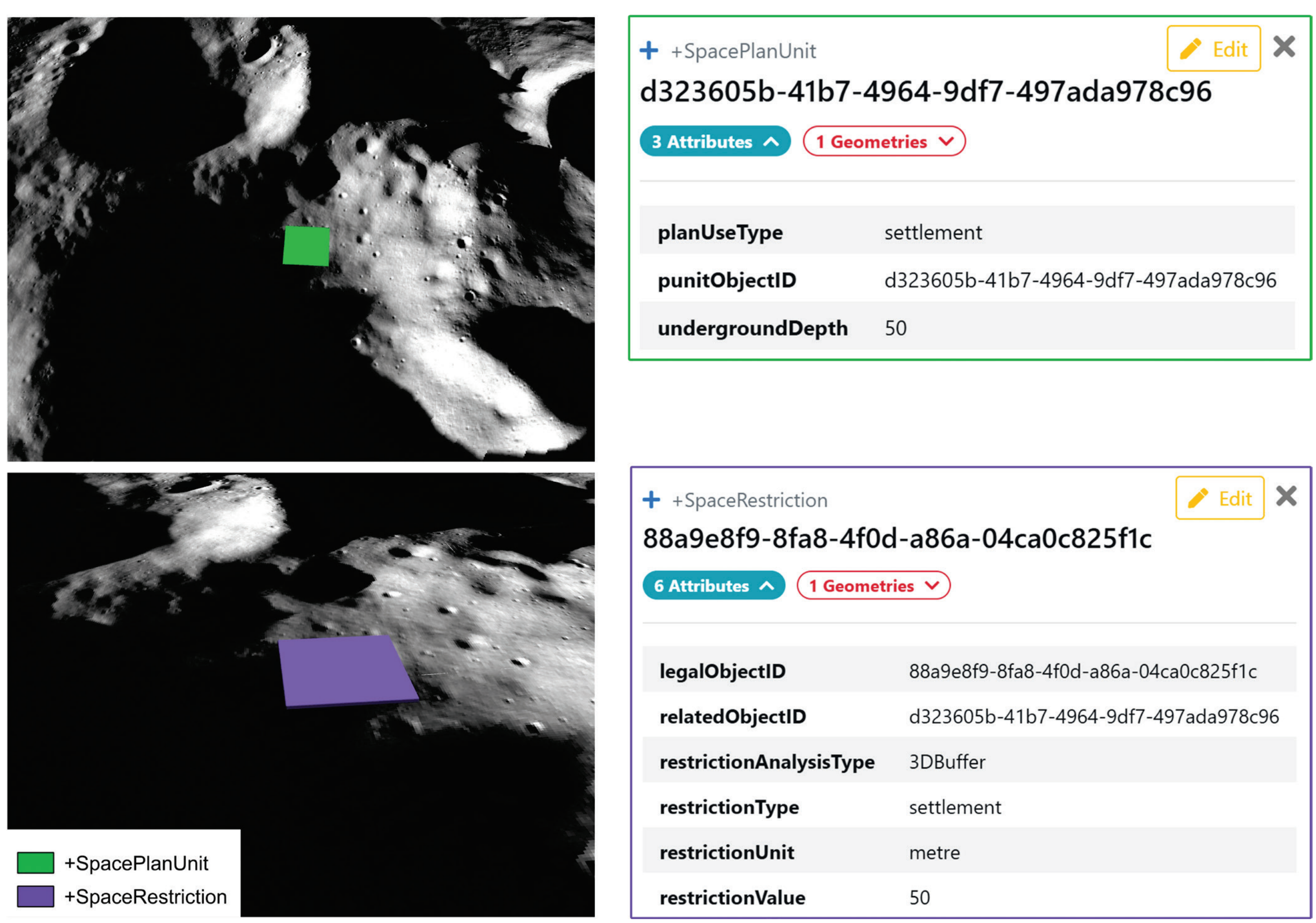

Figure 11. 3D visualization of created +*SpacePlanUnit* and +*SpaceRestriction* objects regarding possible lunar settlement. The border colors of the boxes are matched with the color of the objects.

As shown in Figure 2, different restrictions that are delineated through logical spaces can be defined for +*SpacePlanUnit* objects. In Figure 11, a +*SpaceRestriction* instance is connected with the mentioned +*SpacePlanUnit* instance with the ID "d323605b-41b7-4964-9df7-497ada978c96". As depicted in Figure 11, +*SpaceRestriction* has different attributes that specify the applied restriction. For instance, *restrictionAnalysisType*, *restrictionValue*, and *restrictionUnit* respectively express that 3D Buffer analysis with the value of 50 m is applied. In addition, *relatedObjectID* enables a connection with the related +*SpacePlanUnit* instance.

Figure 12 illustrates the 3D visualization of a +*SpaceRestriction* instance that represents the logical space for the restriction regarding scientific evidence with a +*SpaceScientificEvidence* instance.

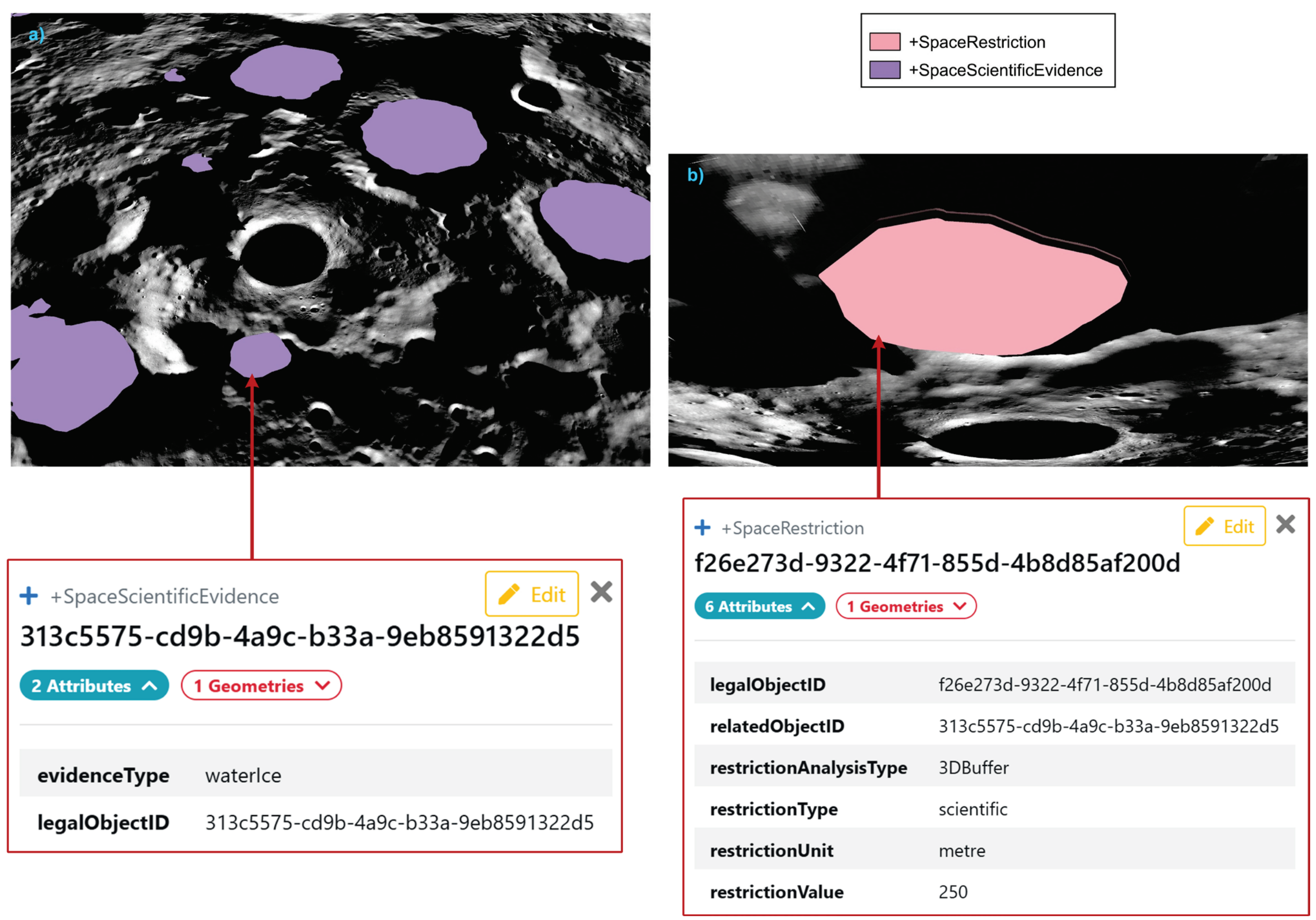

Figure 12. 3D visualization of created *+SpaceScientificEvidence* and *+SpaceRestriction* objects regarding a PSR from surface (a) and subsurface (b) views.

As seen in Figure 12a, the *evidenceType* attribute indicates the type of scientific evidence as water-ice with the value of *waterIce*. This enumeration is in line with the proposed conceptual model in Figure 2. *+SpaceRestriction* instance in Figure 12a indicates the restriction specifications similar to the instance in Figure 11; however, in this case, it has a different value for *restrictionType* attributes as *scientific*. It can also be noted that Figure 12 covers the 3D models of the PSRs in Figure 4 as *+SpaceScientificEvidence* and their related logical spaces as *+SpaceRestriction*. Both object types are modeled based on the modeling details in Table 1 and Table 3.

Figure 13 shows the 3D visualization of *+SpaceProtectedArea* instance that delineates a historical landing site.

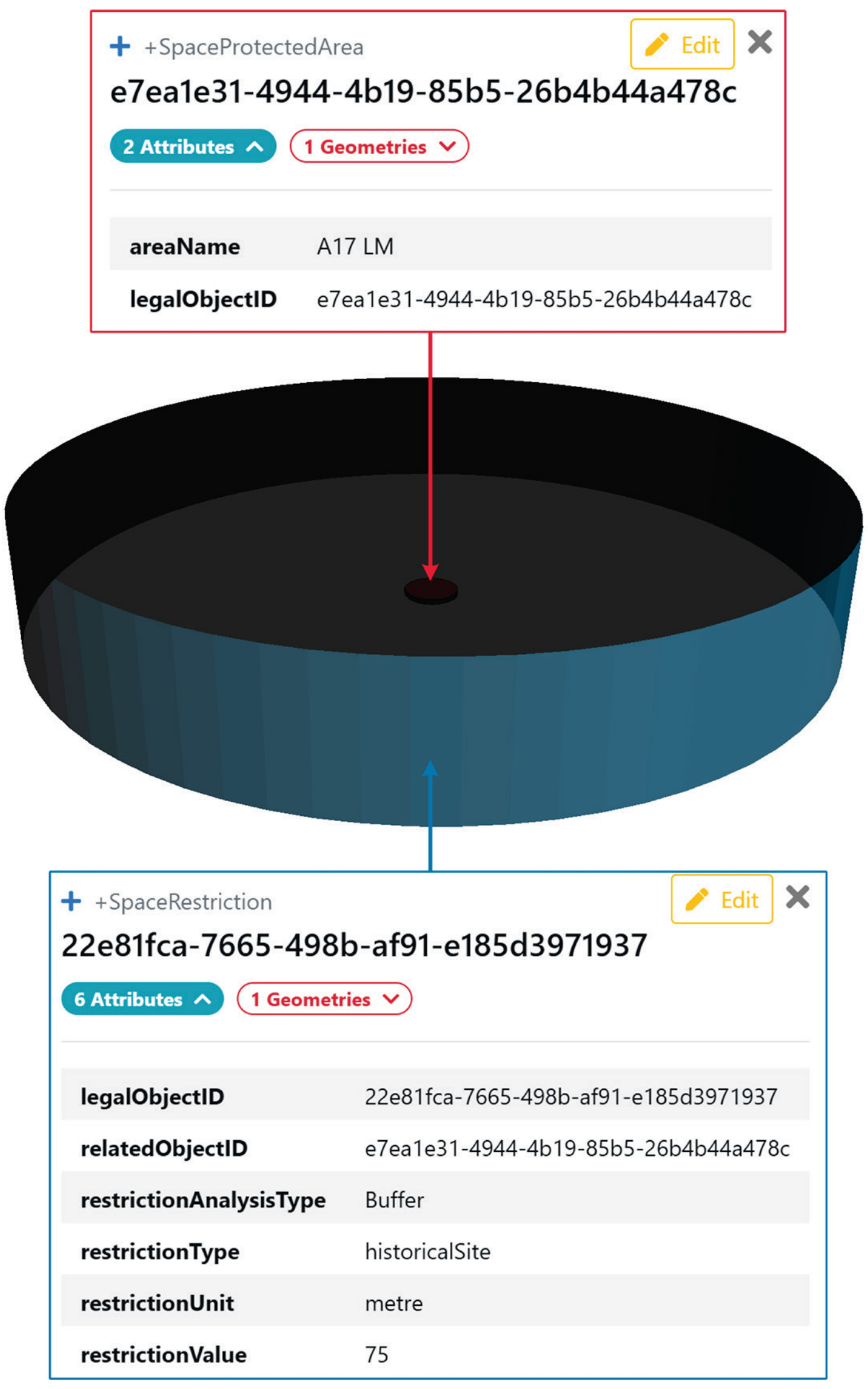

Figure 13. 3D visualization of created +*SpaceProtectedArea* and +*SpaceRestriction* objects regarding historical sites. The border colors of the boxes are matched with the color of the objects. Transparency is applied for +*SpaceRestriction* for visualization purposes.

The specification for this site is indicated through the *areaName* attribute having an A17 LM value. Figure 13 also illustrates the 3D exclusion area that is designated to protect this space heritage site. The type of the applied restriction is defined through the *restrictionType* attribute with the value of *historicalSite*. As can be seen in Figure 13, *restrictionValue* is applied as 75 m. Apart from previous examples, this +*SpaceRestriction* instance is based on a 2D buffer and extrusion rather than a 3D buffer, which can be seen through the value of *restrictionAnalysisType*. Noteworthy to mention is that these specifications match the modeling details in Table 3. Figure 14 shows the 3D visualization of a +*SpaceSurfaceObject* instance that depicts an IP object.

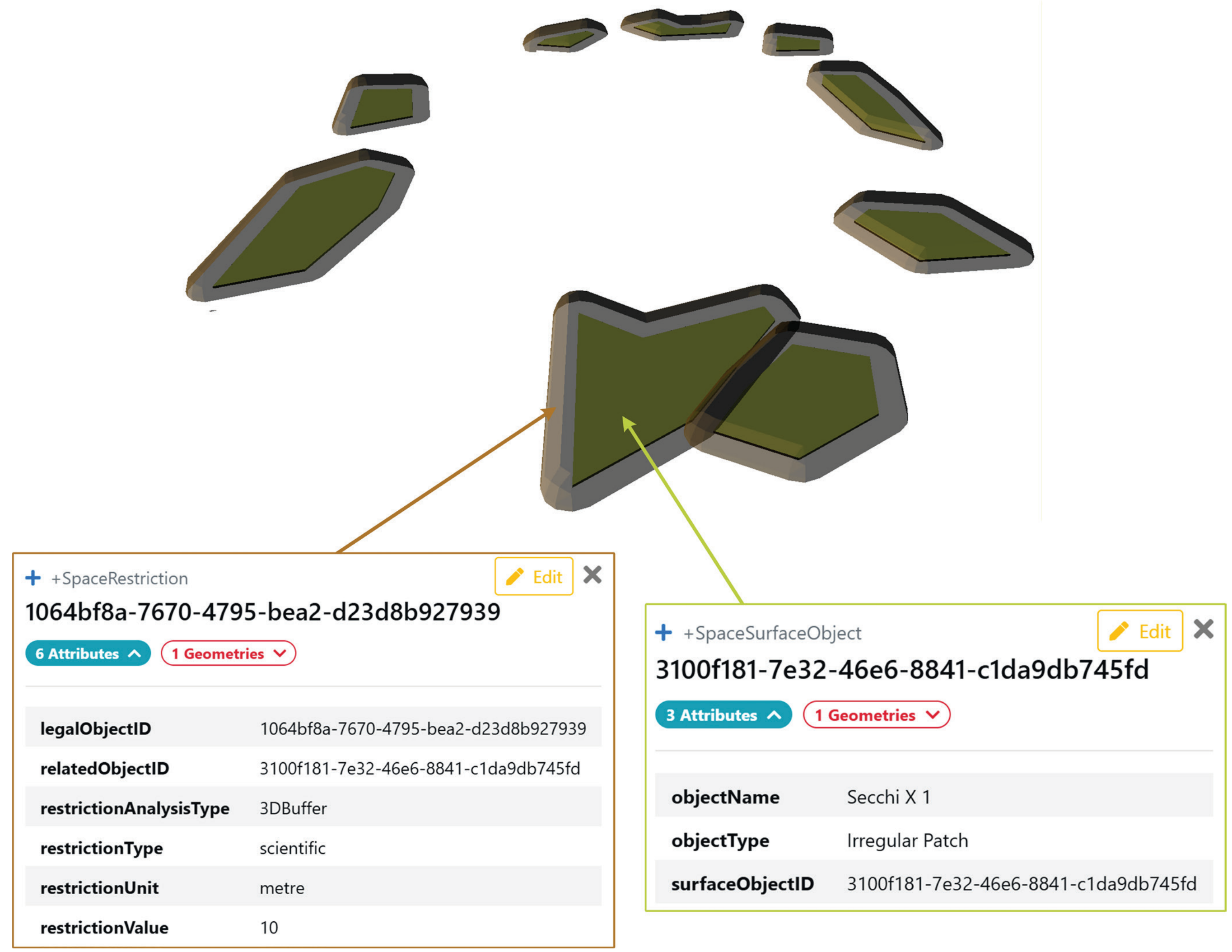

Figure 14. 3D visualization of created +*SpaceSurfaceObject* and +*SpaceRestriction* objects regarding IP. The border colors of the boxes are matched with the color of the objects. Transparency is applied to +*SpaceRestriction* for visualization purposes.

This can be seen through the *objectType* attribute and its value, namely *Irregular Patch*. +*SpaceSurfaceObject* instance also has an *objectName* attribute, which represents the specific name of the modeled IP. Since IP is of scientific interest, different restrictions can be defined. In this connection, Figure 14 illustrates an exemplary restriction by means of a +*SpaceRestriction* instance. As shown in this figure, *restrictionType* attribute of this instance has the value of *scientific* accordingly. It can be mentioned that +*SpaceProtectedArea* and +*SpaceSurfaceObject* instances in Figure 13 and Figure 14 are the selections from historical landing sites and IPs in Figure 5. Figure 15 shows the 3D visualization of +*SpaceBuilding* and +*SpaceBuildingUnit* objects that represent an exemplary building and its building units on the Moon.

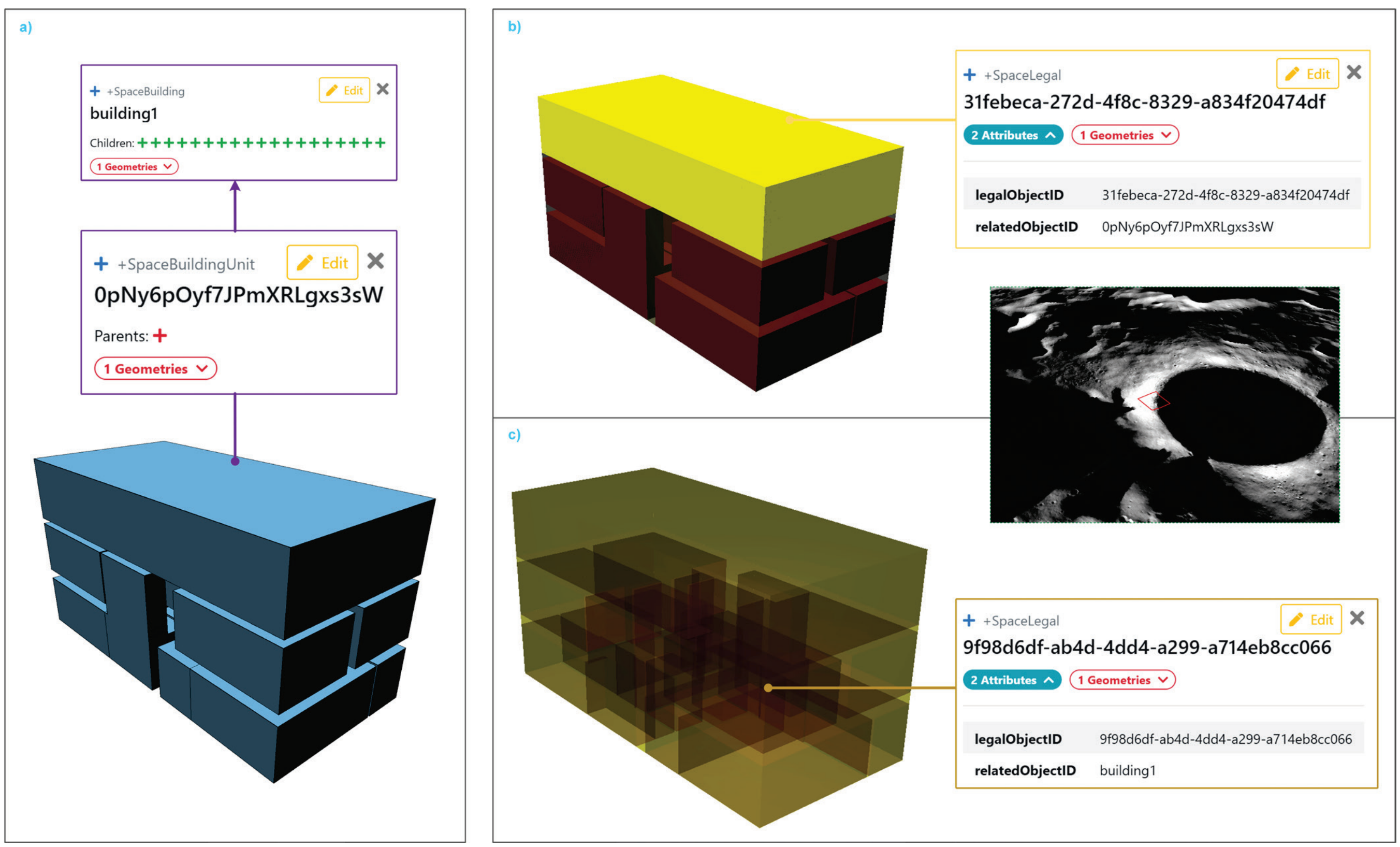

Figure 15. 3D visualization of created +*SpaceBuilding,* +*SpaceBuildingUnit*, and +*SpaceLegal* objects regarding an exemplary building. The transparency is applied for +*SpaceLegal* in Figure 15c for visualization purposes.

These objects are created by means of an FME workbench that transforms the *IfcSpace* instances within the IFC dataset belonging to the building into these objects. Figure 15a illustrates the +*SpaceBuildingUnit* instances within the building, selecting one of these instances with the ID “0pNy6pOyf7JPmXRLgxs3sW”. As outlined before, the proposed extension schema covers the relationship between the feature types. Regarding this, Figure 15a exemplifies the relationship between +*SpaceBuilding* and +*SpaceBuildingUnit* objects, as one of the mentioned relationships. In other words, while the parent object for +*SpaceBuildingUnit* instance with the ID of “0pNy6pOyf7JPmXRLgxs3sW” is defined as +*SpaceBuilding* instance with the ID of “building1”, the children objects, namely the +*SpaceBuildingUnit* instance for this +*SpaceBuilding* instance, is also connected. The relationship between +*SpaceLegal* and +*SpaceBuilding,* as well as +*SpaceBuildingUnit* objects, is established within the proposed extension to enable the 3D modeling of potential logical spaces of the building and building units that can be useful for managing possible exploitation by different parties. Accordingly, Figure 15b illustrates a +*SpaceLegal* instance that is associated with the mentioned +*SpaceBuildingUnit* instance with ID of “0pNy6pOyf7JPmXRLgxs3sW”. Figure 15c depicts the +*SpaceLegal* instance that represents the logical space pertaining to the building.

As mentioned previously in this section, demonstrations on surface objects and related logical spaces on Mars are included in the present study to show the usefulness of the proposed approach for different celestial bodies. In this connection, Figure 16 depicts the +*SpaceCrater* instance that represents the 3D model of Jezero crater on Mars.

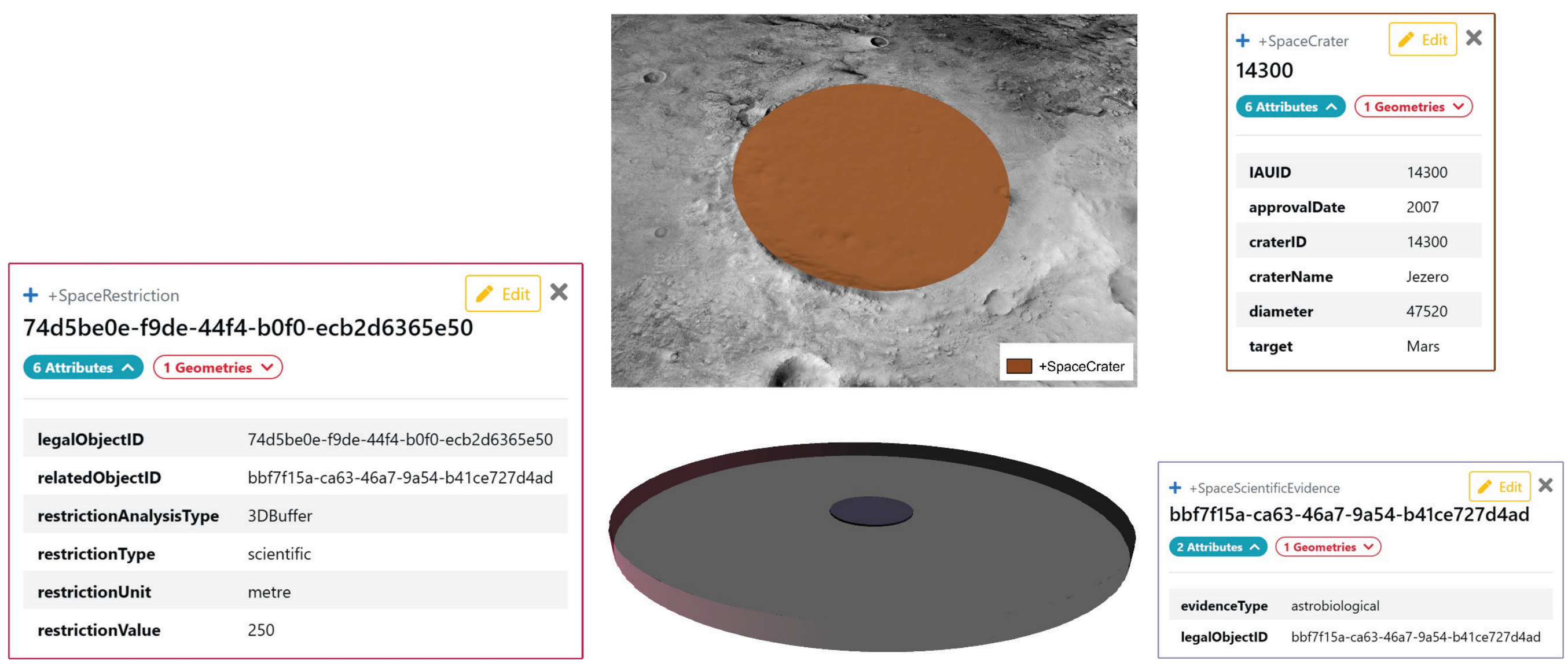

Figure 16. 3D visualization of created +*SpaceCrater,* +*SpaceScientificEvidence*, and +*SpaceRestriction* objects on Mars. The border colors of the boxes are matched with the color of the objects. Transparency is applied to +*SpaceRestriction* for visualization purposes.

This figure also shows the +*SpaceScientificEvidence* instance that delineates the Home Plate region that is of notable scientific interest. As seen in Figure 16, *evidenceType* attribute of this instance has the value *astrobiological* in a way to indicate the potential scientific evidence type. This figure also presents the +*SpaceRestriction* instance that is modeled for delineating the restriction regarding the +*SpaceScientificEvidence* instance. The connection between +*SpaceScientificEvidence* and +*SpaceRestriction* instances can also be seen in Figure 16, similar to the demonstration in Figure 12 that covers 3D objects on the Moon. It can be noted that 3D objects in Figure 16 corresponds to 2D spatial data in Figure 6 that covers Jezero crater and the Home Plate region. Figure 17 presents the selected parts from the content of the CityJSON file that is created for Mars-related objects.

```
{
    "CityObjects": {
        "14300": {
            "attributes": {
                "IAUID": 14300,
                "approvalDate": 2007,
                "craterID": 14300,
                "craterName": "Jezero",
                "diameter": 47520.0,
                "target": "Mars"
            },
            "geometry": [
            "type": "+SpaceCrater"
        },
        "74d5be0e-f9de-44f4-b0f0-ecb2d6365e50": {
            "attributes": {
                "legalObjectID": "74d5be0e-f9de-44f4-b0f0-ecb2d6365e50",
                "relatedObjectID": "bbf7f15a-ca63-46a7-9a54-b41ce727d4ad",
                "restrictionAnalysisType": "3DBuffer",
                "restrictionType": "scientific",
                "restrictionUnit": "metre",
                "restrictionValue": 250
            },
            "geometry": [
            "type": "+SpaceRestriction"
        },
        "bbf7f15a-ca63-46a7-9a54-b41ce727d4ad": {
            "attributes": {
                "evidenceType": "astrobiological",
                "legalObjectID": "bbf7f15a-ca63-46a7-9a54-b41ce727d4ad"
            },
            "geometry": [
            "type": "+SpaceScientificEvidence"
        }
    },
```
a)

```
    "referenceSystem": "https://www.opengis.net/def/crs/EPSG/0/103885"
},
"type": "CityJSON",
"version": "2.0",
```
b)

```
"extensions":
{
    "3DSpace":
    {
        "url": "https://raw.githubusercontent.com/geospatialstudies/space/refs/heads/main/space.ext.json",
        "version": "2.0"
    }
}
```
c)

Figure 17. The selected parts of the created CityJSON file that encompasses surface/subsurface objects and related logical spaces on Mars.

Figure 17a shows the content for *+SpaceCrater*, *+SpaceRestriction*, and *+SpaceScientificEvidence* instances, which belong to instances in Figure 16. It can be underlined that the attributes within Figure 16 match the attributes within the content of Figure 17a. Likewise, the object IDs of these instances, such as “14300”, are the same within the mentioned figures. Figure 17b depicts how the coordinate reference system is defined within the created CityJSON dataset. As mentioned before, CRS 103885 is used for modeling the Mars-related objects. Figure 17c illustrates how the specific extension file is included within the CityJSON dataset. As shown in this figure, the extension file that is developed within this study and shared within the open repository is defined in the created CityJSON dataset. Based on the applied tests, it is important to note that the created CityJSON datasets can be smoothly imported and visualized within QGIS, FME Data Inspector, and *ninja,*[30] which is the web-based official visualization tool for CityJSON (Vitalis, Labetski, et al., 2020). The quadrangles that show the attribute details in the above-mentioned figures are obtained by visualizing the datasets within this tool. They can also be successfully utilized through the most widely used tools related to CityJSON, such as *cjio*, *cjval*, and *val3dity*.

## 6 Discussion and Conclusions

This work provides the CityJSON extension that can be used for creating 3D standardized geodatasets representing the different surface/subsurface objects and their related logical spaces on celestial bodies. It can be mentioned that managing the activities on celestial bodies peacefully is mainly based on suggestions/agreements. For example, the safety zones concept, in which the areas with specific activities, such as mining, are depicted, is included in the Artemis Accords, in order to facilitate efficient planning and executing these activities. In addition, the designated areas concept that suggests the depiction of different areas for different purposes, such as operation and coordination buffer areas, is also proposed for effective management of activities on the lunar surface (Tiballi, 2025). However, even though these concepts include the spatial content contextually, there is a need for clarifying how the information regarding such concepts can be shared in such a way that they are accessible to different stakeholders/parties that aim to carry out activities in similar locations. In this sense, the present study provides a significant contribution to the technical implementation of such concepts because it enables the modeling of the defined areas as standardized geodatasets that can be stored and shared in an interoperable manner. Another related contribution of this research is that even though the depiction of these specific areas was mentioned as 2D-based analysis, such as a 0.1 km buffer in the previous studies, they can be formed as 3D geodatasets that are modeled based on 3D-based analysis, such as 3D buffers, as demonstrated in Figure 11.

This paper provides significant results that can be used as a technical foundation for planning space mining activities (see Figure 10). First, it is shown how the mining locations on celestial bodies such as the Moon can be modeled in a standardized manner. Second, the current research presents how the logical spaces that delineate the areas where the mining activities can be conducted by which party/parties can be modeled in 3D by considering the underground.

---

[30] https://ninja.cityjson.org/#

Third, it is evidenced that previously proposed concepts regarding planning mining activities, for example, the one by Hubbard et al. (2024), can be enhanced in a way to consider the 3D through interoperable geodatasets. The present results can also be useful for creating legal frameworks on space-based activities.

The current study also provides crucial results that can contribute to the discussions on property rights regarding celestial bodies, since it illustrates how the logical spaces that might be considered as legal space as well can be represented by 3D standardized geodatasets. This is important because the technical implementation of legal frameworks might be inefficient if such a framework is prepared without considering the current technical approaches and capabilities. Land administration practices on Earth can be given as an example of this situation. The legal frameworks on which these practices are based need to be reorganized such that they can be readily implemented through current techniques, such as BIM, in order to meet the requirements due to the increased complexity of the built environment (Sun et al., 2023). In particular, 3D should be considered when planning and managing the underground space in urban areas (Guler, 2024). Similar consideration for space-based activities can be a topic of the planned legal frameworks regarding celestial bodies.

Another related contribution regarding the abovementioned issue is that different logical spaces that depict the restrictions can be modeled comprehensively by means of the proposed approach. This is because different types of these restrictions, such as those related to historical sites, are also within the topic of land administration practices on Earth. Such restrictions are commonly referred to as cadastral restrictions since they can be depicted through physical objects with spatial content. For example, the mentioned restrictions in the below of the land require significant considerations when planning activities on the underground space, such as designing the underground metro tunnel. Moreover, different restrictions can be defined for the areas that are of important scientific interest, such as water-ice evidence, as demonstrated in Figure 12 and Figure 14. Accordingly, the abovementioned restrictions that encompass the above and below of the surface of celestial bodies can be included in the considerations regarding the underground activities, such as designing the subsurface lunar bases and mining activities.

Researchers mentioned in the literature that 2D restrictions might be insufficient to efficiently protect the highly valuable areas on celestial bodies, such as astrobiological evidence and heritage sites, from external factors, such damages due to exhaust-driven rocket engine (Fletcher et al., 2024; Spennemann & Murphy, 2020). The present work, therefore, provides a quite vital contribution to the literature considering the mentioned issue. This results from the fact that it not only enables the standardized geodatasets that delineate the specific areas, such as scientific evidence and heritage sites, but also paves the way for representing logical spaces in an interoperable manner where 3D restrictions/exclusions related to these areas are applied. It is important to note that multiple restrictions that depict the zones with different quantities, such as 1 km and 2 km, can be defined for a specific area of interest by means of the proposed approach.

This research also provides a key foundation that can make a contribution to efforts regarding the standardization of planetary-based spatial datasets. The proposed conceptual model that covers the different surface/subsurface objects and their related logical spaces can be used for developing the conceptual models within PSDI. This is one of the fundamental components for establishing the PSDI, similar to SDI in a terrestrial context. Furthermore, these conceptual models should indicate geometry specifications of the feature types that represent the physical objects from different contexts, such as geology. Additionally, there is a current interest in improving the terrestrial SDI such that it can utilize 3D spatial datasets, so as to holistically manage the multifaceted counterparts of the urban environment, such as underground space. In relation to this, it can be noted that previous studies didn’t focus on 3D in the context of the PSDI. The current study thus contributes to establishing a PSDI that includes three dimensions because it demonstrates how 3D geodatasets that are validated against the proposed conceptual model can be created by considering the feature modeling requirements regarding surfaces of the celestial bodies. As a related contribution, this work extends the use of a 3D geoinformation-based standard beyond its original Earth-based built-environment focus to a celestial body context.

Several issues can be pointed out in terms of technical implementation. First, it can be noted that the standardization regarding the coordinate reference systems is highly important to be able to create interpretable geodatasets. This issue was clearly highlighted in the literature (Archinal et al., 2020; Hare et al., 2018; Paganelli et al., 2021). Using a pre-defined CRS is a requirement for creating a CityJSON dataset since it should be indicated with its code within such a dataset. As mentioned in Section 5, the specific CRSs, such as 103878 (Moon) and 103885 (Mars) are utilized in this study in the context of data management and transformation.

Using them allows for organizing data in 2D within QGIS and applying spatial analysis and transforming data within FME. In addition, the created CityJSON datasets can be readily imported and amended within the tools in the sense of coordinate system compatibility since they contain the pre-defined CRS. Sharing spatial datasets that show the results of scientific research with a pre-defined CRS should be promoted strongly in order to increase the reusability and consequently the benefits of these datasets. Second, it can be mentioned to the spatial resolution of the DEM that is used to create 3D objects that represent the craters. Two of the selected craters on the Moon, namely Shoemaker and Shackleton, have highly large diameters, approx 52 km and 21 km, respectively. When the high-resolution DEM (i.e., 5 m) was used, computing memory was insufficient to create the 3D objects with this data source. For this reason, the existing DEM was aggregated into 100 m resolution, and hence 3D objects of the craters were able to be created. In the case of Jezero crater on Mars, a 3D object was formed using a DEM with 50 m resolution, even though this crater has approx. 48 km in diameter, which is similar to the selected craters on the Moon. This 3D object was created by using a DEM having an increased spatial resolution, and the possible reason for this is that the selected craters on the Moon have more variation in terms of depth. This is due to the fact that creating multisurface geometry based on the point clouds requires much more memory resources when the geometry is more complex. In this regard, the selection of the source DEM can be examined depending on the requirements within the case where *+SpaceCrater* instances will be exploited. Third, the values in Table 1 and Table 3 that are used to create the 3D models of surface/subsurface objects and logical spaces are selected for demonstration purposes. It can be underlined that these values can be determined based on the consensus within the frameworks for different contexts, such as space heritage. Accordingly, 3D models can be formed based on the determined values by means of the proposed approach in the current study.

Noteworthy to mention is that the present study aims to extend the core schema of the CityJSON standard rather than developing a new exchange format for several reasons. For example, this standard is developed in a way to allow for extending its core schema based on the data requirements for different disciplines and application areas. Therefore, the developers of the CityJSON significantly intend to enable effortless usability of datasets that are produced according to the proposed extensions without a need for any extra adjustments within the software and tools. This issue is also an important reason for not proposing a new exchange format in the present study. This is because when a new exchange format is proposed, there is a need for modifying the existing tools and software such that they support that format. Otherwise, the efficient implementation and usability of the produced datasets based on this new exchange format would be hindered. Moreover, CityJSON provides a wide range of open-source tools for supporting validation, updating, and visualization of the datasets that are created in line with an extension, which their efficient usability is exercised in the present study.

It is notable that the current work might have a limitation regarding the data conversion that produces CityJSON datasets. This is because FME might not be utilized by every scholar since it is not an open tool, even though it might be exploited freely through its grant program, which is the strategy followed in this study. FME is used in this research because it is widely used for spatial data transformation and can write CityJSON datasets. However, it is worth noting that one of the most convenient ways to achieve an open-source approach is to develop a Python-based script that transforms 2D spatial datasets (e.g., GeoJSON) into 3D CityJSON datasets. *3dfier*[31], an open-source tool for creating 3D models including CityJSON datasets by using 2D and point cloud datasets, can be given as an example in this sense (Ledoux et al., 2021). According to the standard's official website, there are only three tools capable of generating CityJSON datasets; notably, *3dfier* is the only one that is open-source[32]. Furthermore, as mentioned above, after obtaining the CityJSON datasets, the validation of them against the proposed extension can be realized easily by means of the provided open-source tools.

Some points can be mentioned to provide a basis for future studies. First, the database implementation was not included in this work since its main aim is to show how 3D standardized geodatasets that represent the surface/subsurface objects and logical spaces can be created. Therefore, suitable ways for enabling the storage and hence distribution of these 3D geodatasets can be examined. This is also essential for designing the PSDI in a way to take the 3D geometries into consideration. Second, as complementary to the previous issue, the recent standardization, such as the OGC Application Programming Interface (API)[33] regarding sharing geospatial datasets over the internet, can be investigated for serving the created 3D models in the present research in order to facilitate the widespread adoption and usefulness of these models. Third, the content and applicability of the proposed conceptual model can be further analyzed based on the opinions of the experts from different disciplines. The proposed approach in this study can be replicated even if this model is modified. The feature types and attributes are designed in a foundational manner so that they could be enhanced for different celestial bodies, including asteroids. One example of this enhancement is the consideration of subsurface extraction in space mining applications. It is also notable that future studies can focus on developing a more neutral conceptual model without specifically considering CityJSON in order to provide a basis for other implementations.

---

[31] https://github.com/tudelft3d/3dfier

[32] https://www.cityjson.org/software/

[33] https://ogcapi.ogc.org/

In relation to this issue, the possibility of integrating celestial body-based adjustment/implementation of LADM that aims to provide standardization for land registry and cadastre processes on Earth can be investigated in future studies. In the present study, a more compact conceptual model that deals with surface/subsurface objects and related logical spaces for sustainable space exploration is provided since there does not exist clear consensus and descriptions regarding property rights and registration in the context of celestial bodies. It is worth noting that there is an ongoing development for improving the implementation processes of the LADM conceptual model in collaboration with international organizations, including OGC (van Oosterom et al., 2025). In this sense, the current study can be exploited as a foundational work for demonstrating the possible usability of LADM within the space/celestial body context since it provided successful implementation of the developed conceptual model through a 3D spatial data standard.

Lastly, it can be mentioned that this research significantly focuses on enabling 3D geodatasets representing the surface/surface objects and their related logical spaces in a standardized way through developing a foundational conceptual model. However, future studies can investigate the visualization of the geodatasets that will be produced in line with the developed model using Cesium Moon[6] and Mars[7] datasets that enable complete 3D Tiles of the surfaces of these celestial bodies.

**Conflict of Interest statement**

None of the authors has a conflict of interest to disclose.

**Acknowledgments**

We acknowledge the support of the Canadian Space Agency (CSA) [25EXPROSS4]. The ideas expressed are those of the authors only and do not represent the views of affiliated/supporting organizations. D.C.G. acknowledges the Lunar Meteoroid Impact Observer (LUMIO) Mission Science Team as a member.

## Data Availability

The data related to this study are openly available at https://github.com/geospatialstudies/space. The sources of all input data are cited in the article.